\documentclass[10pt,twocolumn]{IEEEtran}

\usepackage{amsfonts,amssymb,amsmath}

\newtheorem{Def}{Definition}
\newtheorem{Lem}{Lemma}
\newtheorem{Thrm}{Theorem}

\begin{document}
\title{Unconditionally Secure Key Distribution In Higher Dimensions By
Depolarization}

\author{H. F. Chau, Member, IEEE\thanks{H. F. Chau is with the Department of
 Physics, University of Hong Kong, Pokfulam Road, Hong Kong. (E-mail:
 hfchau@hkusua.hku.hk)}}
\maketitle

\begin{abstract}
 This paper presents a prepare-and-measure scheme using $N$-dimensional quantum
 particles as information carriers where $N$ is a prime power. One of the key
 ingredients used to resist eavesdropping in this scheme is to depolarize all
 Pauli errors introduced to the quantum information carriers. Using the
 Shor-Preskill-type argument, we prove that this scheme is unconditionally
 secure against all attacks allowed by the laws of quantum physics. For $N =
 2^n > 2$, each information carrier can be replaced by $n$ entangled qubits. In
 this case, there is a family of eavesdropping attacks on which no
 unentangled-qubit-based prepare-and-measure quantum key distribution scheme
 known to date can generate a provably secure key. In contrast, under the same
 family of attacks, our entangled-qubit-based scheme remains secure whenever
 $2^n \geq 4$. This demonstrates the advantage of using entangled particles as
 information carriers and of using depolarization of Pauli errors to combat
 eavesdropping attacks more drastic than those that can be handled by
 unentangled-qubit-based prepare-and-measure schemes.
\end{abstract}


\section{Introduction \label{Sec:Intro}}
\PARstart{K}{ey} distribution is the art of sharing a secret key between two
 cooperative players Alice and Bob in the presence of an eavesdropper Eve. If
 Alice and Bob distribute their key by exchanging classical messages only, Eve
 may at least in principle wiretap their conversations without being caught.
 So, given unlimited computational resources, Eve can crack the secret key. In
 contrast, in any attempt to distinguish between two non-orthogonal states,
 information gain is only possible at the expense of disturbing the state
 \cite{bbjm}. Therefore, if Alice and Bob distribute their secret key by
 sending non-orthogonal quantum signals, any eavesdropping attempt will almost
 surely affect their signal fidelity. Consequently, a carefully designed
 quantum key distribution (QKD) scheme allows Alice and Bob to accurately
 determine the quantum error rate, which in turn reflects the eavesdropping
 rate. If the estimated quantum error rate is too high, Alice and Bob abort the
 scheme and start all over again. Otherwise, they perform certain privacy
 amplification procedures to distill out the final key
 \cite{biasedbb84,lochauqkdsec,mayersjacm,gottloreview,gisinreview}. It is,
 therefore, conceivable that a provably secure QKD scheme exists even when Eve
 has unlimited computational power.

 With this belief in mind, researchers proposed many QKD schemes
 \cite{gisinreview}. These schemes differ in many ways such as the Hilbert
 space dimension of the quantum particles used, as well as the states and bases
 Alice and Bob prepared and measured. The first QKD scheme, commonly known as
 BB84, was invented by Bennett and Brassard \cite{bb84}. In BB84, Alice
 randomly and independently prepares each qubit in one of the following four
 states: $|0\rangle$, $|1\rangle$ and $(|0\rangle \pm |1\rangle)/\sqrt{2}$, and
 sends them to Bob. After receiving the qubits, Bob randomly and independently
 measures each qubit in either $\{ |0\rangle, |1\rangle \}$ or $\{ (|0\rangle
 \pm |1\rangle)/\sqrt{2} \}$ bases. In short, BB84 is an experimentally
 feasible prepare-and-measure (PM) scheme involving the transfer of unentangled
 qubits \cite{bb84}. Later, Bru{\ss} introduced another experimentally feasible
 PM scheme known as the six-state scheme \cite{sixstate}. In this scheme, Alice
 randomly and independently prepares each qubit in one of the following six
 states: $|0\rangle$, $|1\rangle$, $(|0\rangle \pm |1\rangle)/\sqrt{2}$ and
 $(|0\rangle \pm i |1\rangle)/\sqrt{2}$; and Bob measures each of them randomly
 and independently in one of the following three bases: $\{ |0\rangle, |1
 \rangle \}$, $\{ (|0\rangle \pm |1\rangle)/\sqrt{2} \}$ and $\{ (|0\rangle \pm
 i|1\rangle)/\sqrt{2} \}$. Although the six-state scheme is more complex and
 generates a key less efficiently, Bru{\ss} found that it tolerates higher
 noise level than BB84 if Eve attacks each qubit individually \cite{sixstate}.
 In addition to qubit-based schemes such as BB84 and the six-state scheme, a
 number of PM QKD schemes involving higher dimensional or continuous systems
 have been proposed \cite{cont1,cont2,squeez,threestate,alph,highdim,dlevel,
 highdimmore,earlier_version}. Most importantly, compared with qubit-based PM
 schemes, studies showed that many PM schemes involving higher dimensional
 systems can generate secure keys when a higher fraction of particles is
 eavesdropped individually
 \cite{alph,highdim,dlevel,highdimmore,evethreestate}.

 Instead of using PM schemes, Alice and Bob may explicitly use their shared
 entanglement to create a secret key. The first such entanglement-based (EB)
 QKD scheme was proposed by Ekert \cite{Ekert91}. This scheme makes use of the
 fact that measuring a singlet state $(|01\rangle - |10\rangle)/\sqrt{2}$ along
 a common axis produces a pair of anti-correlated random bits. Consequently, a
 common key can be established provided that Alice and Bob share singlets
 through a quantum communication channel. To ensure that the fidelity of the
 shared singlets is high, Alice and Bob check if certain Bell's inequalities
 are maximally violated in a randomly selected subset of their shared particles
 \cite{Ekert91}. Comparing with PM schemes, a typical EB scheme generates a key
 more efficiently but is harder to implement experimentally.

 Are these QKD schemes really secure? Is it true that the six-state scheme
 tolerates higher error level than BB84? The answers to these questions turn
 out to be highly non-trivial. Recall that the all powerful Eve may choose to
 attack the transmitted qubits collectively by applying a unitary operator to
 entangle these qubits with her quantum particles. In this situation most of
 our familiar tools such as classical probability theory do not apply to the
 resultant highly entangled non-classical state. These make rigorous
 cryptanalysis of BB84, the six-state and Ekert schemes extremely difficult.

 In spite of these difficulties, air-tight security proofs against all possible
 eavesdropping attacks of BB84, the six-state and Ekert schemes have been
 discovered. Rigorous proofs of QKD schemes with better error tolerance have
 also been found. Mayers \cite{mayersjacm} and Biham \emph{et al.} \cite{biham}
 eventually proved the security of BB84 against all kinds of attacks allowed by
 the known laws of quantum physics. In particular, Mayers showed that in BB84 a
 provably secure key can be generated whenever the bit error rate is less than
 about 7\% \cite{mayersjacm}. (A precise definition of bit error rate can be
 found in Def.~\ref{Def:Error_Rates} in Subsection~\ref{Subsec:Error_Rates}.
 Moreover, we emphasize that, unless otherwise stated, all provably secure
 error rates quoted in this paper are provable lower bounds. A QKD scheme may
 generate a secure key at a higher error rate although a rigorous proof has not
 been found.) Along a different line, Lo and Chau \cite{lochauqkdsec} proved
 the security of an EB QKD scheme, which is similar to the Ekert scheme, that
 applies up to 1/3 bit error rate by means of a random hashing technique based
 on entanglement purification \cite{bdsw}. Their security proof is conceptually
 simple and appealing. Nevertheless, their scheme requires quantum computers
 and hence is not practical yet. By ingeniously combining the essence of the
 Mayers and Lo-Chau proofs, Shor and Preskill gave a security proof of BB84
 that applies up to 11.0\% bit error rate \cite{shorpre}. This is a marked
 improvement over the 7\% bit error tolerance rate in Mayers' proof. Since
 then, the Shor-Preskill proof became a blueprint for the cryptanalysis of many
 QKD schemes. For instance, Lo \cite{sixstateproof} as well as Gottesman and Lo
 \cite{qkd2waylocc} extended it to cover the six-state QKD scheme. At the same
 time, the work of Gottesman and Lo also demonstrates that careful use of local
 quantum operation plus two-way classical communication (LOCC2) increases the
 error tolerance rate of QKD. Furthermore, they found that the six-state scheme
 tolerates a higher bit error rate than BB84 because the six-state scheme gives
 better estimates for the three Pauli error rates \cite{qkd2waylocc}. In search
 of an unentangled-qubit-based (UQB) QKD scheme that tolerates higher bit error
 rate, Chau recently discovered an adaptive entanglement purification procedure
 inspired by the technique used by Gottesman and Lo in Ref.~\cite{qkd2waylocc}.
 He further gave a Shor-Preskill-based proof showing that this adaptive
 entanglement purification procedure allows the six-state scheme to generate a
 provably secure key up to a bit error rate of $(5-\sqrt{5})/10 \approx 27.6\%$
 \cite{sixstateexact}, making it the most error-tolerant PM scheme involving
 the transfer of unentangled qubits to date.
 
 Unlike various UQB QKD schemes, very little cryptanalysis against the most
 general type of eavesdropping attack on a QKD scheme involving the transfer of
 higher dimensional quantum systems or entangled qubits has been performed. The
 only relevant work to date seems to be the earlier version of this work
 \cite{earlier_version}. In that manuscript, an unconditionally secure QKD
 scheme that generalized the six-state scheme by using conjugation to
 cyclically permute $\mbox{O} (N)$ kinds of quantum errors that can occur in
 the $N$-dimensional quantum information carriers was reported. Moreover, the
 set of preparation and measurement bases used is mutually unbiased
 \cite{earlier_version}. Probably because Pauli errors are not depolarized when
 $N>2$, the error tolerance capability of that scheme is not particularly high
 under the most general type of attack when $2 < N \lesssim 16$. More
 importantly, that scheme does not conclusively demonstrate the superiority of
 using entangled qubits to combat Eve \cite{earlier_version}. In contrast,
 almost all cryptanalysis suggests that QKD schemes involving higher
 dimensional systems are more error-tolerant under individual particle attack
 \cite{alph,highdim,dlevel,evethreestate}. It is, therefore, instructive to
 find an unconditionally secure PM QKD scheme based on entangled qubits that
 stands up to more drastic eavesdropping attacks than all known UQB PM schemes
 known to date.

 In this paper, we analyze the security and error tolerance capability of a PM
 QKD scheme involving the transmission of higher dimensional quantum particles
 or entangled qubits. In fact, this scheme makes use of $N$-dimensional quantum
 information carriers prepared and measured randomly in $N(N+1)$ different
 bases. (In the cases of $N=2,3,5,7,11$, the number of bases used can be
 reduced to $(N+1)$.) Such a preparation and measurement procedure depolarizes
 all Pauli errors in the transmitted signal. This greatly restricts the form of
 errors occurring in the quantum signals and makes error estimation effective;
 hence, its error tolerance rate is high. Nonetheless, the high error tolerance
 rate comes with a price, namely, that the efficiency of the scheme drops.

 This paper is organized as follows: We first review the general assumptions on
 the capabilities of Alice, Bob and Eve, as well as a precisely stated security
 requirement for a general QKD scheme in Section~\ref{Sec:Sec_Cond}. Then we
 introduce an EB QKD scheme involving the transmission of $N$-dimensional
 quantum systems, where $N$ is a prime power, in Section~\ref{Sec:EntScheme}
 and prove its security against the most general eavesdropping attack in
 Section~\ref{Sec:SecEntScheme}. By standard Shor and Preskill reduction
 argument, we arrive at the provably secure PM scheme using unentangled
 $N$-dimensional quantum particles in Section~\ref{Sec:ShorPre}. Since one may
 use $n$ possibly entangled qubits to represent an $N$-dimensional quantum
 state whenever $N=2^n$, we obtain an unconditionally secure
 entangled-qubit-based (EQB) PM QKD scheme. (See Section~\ref{Sec:ShorPre} for
 a discussion of a subtle point in constructing this EQB PM QKD scheme.
 Moreover, we emphasize that the term EQB means that the qubits used to
 transfer information between Alice and Bob are entangled. In contrast, the
 term EB means that entanglement shared between Alice and Bob is explicitly
 used to generate the secret key. Thus, an EQB scheme may not be an EB scheme.)
 This EQB PM QKD scheme offers a definite advantage over all currently known
 UQB ones used to combat Eve. Specifically, whenever the most error-tolerant
 UQB PM QKD scheme known to date (namely, the one introduced by Chau in
 Ref.~\cite{sixstateexact}) can generate a provably secure key under an
 eavesdropping attack, this EQB scheme can also generate an equally secure key
 for any $2^n\geq 2$ under the same attack. Furthermore, there is a family of
 eavesdropping attacks that creates a bit error rate too high for Chau's scheme
 in Ref.~\cite{sixstateexact} to generate a provably secure key. In contrast,
 the same family of attacks does not prevent this EQB PM scheme from producing
 a secure key whenever $2^n \geq 4$. This observation convincingly demonstrates
 that using entangled particles as information carriers can increase error
 tolerance in QKD.  Lastly, we give a brief summary in Section~\ref{Sec:Dis}.

\section{General Features And Security Requirements For Quantum Key
 Distribution \label{Sec:Sec_Cond}}
 In QKD, we assume that Alice and Bob have access to two communication
 channels. The first one is an insecure noisy quantum channel. The other one is
 an unjammable noiseless authenticated classical channel in which everyone,
 including Eve, can listen to, but cannot alter, the content passing through
 it. We also assume that Alice and Bob have complete control over their own
 apparatus. Everything else for the unjammable classical channel may be
 manipulated by the all powerful Eve. We further make the most pessimistic
 assumption that Eve is capable of performing any operation in her controlled
 territory that is allowed by the known laws of quantum physics
 \cite{gottloreview,gisinreview}.

 Given an unjammable classical channel and an insecure quantum channel, a QKD
 scheme consists of three stages \cite{biasedbb84}. The first is the signal
 preparation and transmission stage in which quantum signals are prepared and
 exchanged between Alice and Bob. The second is the signal quality test stage
 in which a subset of the exchanged quantum signals is measured in order to
 estimate the eavesdropping rate in the quantum channel. The final phase is the
 signal privacy amplification stage in which a carefully designed privacy
 amplification procedure is performed to distill out an almost perfectly secure
 key.

 No QKD scheme can be 100\% secure as Eve may be lucky enough to guess the
 preparation or measurement bases for each quantum state correctly. Hence, it
 is more reasonable to demand that the mutual information between Eve's
 measurement results after eavesdropping and the final secret key is less than
 an arbitrary but fixed small positive number. Hence we adopt the following
 definition of security.

\par\medskip
\begin{Def}[Based on Lo and Chau \cite{lochauqkdsec}]
 With the above assumptions on the unlimited computational power of Eve, a QKD
 scheme is said to be \textbf{unconditionally secure} with security parameters
 $(\epsilon_p,\epsilon_I)$ provided that whenever Eve has a cheating strategy
 that passes the signal quality control test with probability greater than
 $\epsilon_p$, the mutual information between Eve's measurement results from
 eavesdropping and the final secret key is less than $\epsilon_I$.
 \label{Def:Security}
\end{Def}

\section{An Entanglement-Based Quantum Key Distribution Scheme
\label{Sec:EntScheme}}
 In this section, we generalize the six-state scheme in a new way. In
 Subsection~\ref{Subsec:TConstruct}, we first identify each element in
 $SL(2,N)$, the special linear group of $2\times 2$ matrices over the finite
 field $GF(N)$, with a distinct unitary operator in $U(N)$. It turns out that
 all Pauli errors occurring in the transmitted particles can be depolarized by
 conjugating each transmitted particle by a randomly and independently picked
 unitary operator to be constructed. Then, in
 Subsection~\ref{Subsec:EntScheme}, we devise an EB QKD scheme based on this
 set of unitary operators.

\subsection{Construction Of The Unitary Operator $T(M)$
\label{Subsec:TConstruct}}
 We begin with the following definitions.

\par\medskip
\begin{Def}[Ashikhmin and Knill \cite{knill}]
 Let $a\in GF(N)$ where $N = p^n$ with $p$ being a prime. We define the unitary
 operators $X_a$ and $Z_a$ acting on an $N$-dimensional Hilbert space by
 \begin{equation}
  X_a |b\rangle = |a+b\rangle \label{E:DefX_a}
 \end{equation}
 and
 \begin{equation}
  Z_a |b\rangle = \chi_a (b) |b\rangle \equiv \omega_p^{\textrm{Tr} (a b)}
  |b\rangle , \label{E:DefZ_a}
 \end{equation}
 where $\chi_a$ is an additive character of the finite field $GF(N)$,
 $\omega_p$ is a primitive $p$th root of unity and $\textrm{Tr} (a) = a + a^p +
 a^{p^2} + \cdots + a^{p^{n-1}}$ is the absolute trace of $a\in GF(N)$. Note
 that the arithmetic inside the state ket and in the exponent of $\omega_p$ is
 performed in the finite field $GF(N)$. \label{Def:X_a_and_Z_a}
\end{Def}

\par\medskip
 It is easy to see from Definition~\ref{Def:X_a_and_Z_a} that the set of all
 Pauli errors acting on an $N$-dimensional particle $\{ X_a Z_b : a,b \in GF(N)
 \}$ spans the set of all possible linear operators acting on that particle
 over ${\mathbb C}$. (Unless otherwise stated, all linear operators discussed
 in this paper are endomorphisms.) Besides, $X_a$ and $Z_b$ follow the algebra
\begin{equation}
 X_a X_b = X_b X_a = X_{a+b} , \label{E:Algebra_X_X}
\end{equation}
\begin{equation}
 Z_a Z_b = Z_b Z_a = Z_{a+b} \label{E:Algebra_Z_Z}
\end{equation}
 and
\begin{equation}
 Z_b X_a = \omega_p^{\textrm{Tr} (a b)} X_a Z_b \label{E:Algebra_X_Z}
\end{equation}
 for all $a,b\in GF(N)$, where arithmetic in the subscripts is performed in
 $GF(N)$.

 One way to permute quantum errors is to construct a unitary operator that maps
 $X_a Z_b$ to $X_{a\alpha+b\beta} Z_{a\delta+b\gamma}$ modulo a phase factor by
 conjugation. Specifically, let $M = \left[ \begin{array}{cc} \alpha & \beta \\
 \delta & \gamma \end{array} \right] \in SL(2,N)$ where $N = p^n$ is a prime
 power. We look for a unitary operator $T(M)$ satisfying
\begin{equation}
 T(M)^{-1} X_a Z_b \,T(M) = \omega_p^{f_M(a,b)} X_{a\alpha + b\beta} Z_{a\delta
 + b\gamma} \label{E:XZT}
\end{equation}
 for all $a,b\in GF(N)$, where the arithmetic in the subscripts is performed in
 $GF(N)$ and the factor $\omega_p^{f_M(a,b)} \in {\mathbb C}$ satisfies
 $|\omega_p^{f_M(a,b)}| = 1$. When the matrix $M\in SL(2,N)$ is clearly known
 to the readers, we shall simply denote $T(M)$ by $T$ and $f_M$ by $f$.

 The choice of $T$ is not unique if it exists. This is because $e^{i\theta} X_c
 Z_d T$ also permutes quantum errors modulo a phase factor for all $\theta\in
 {\mathbb R}$ and $c,d\in GF(N)$. (However, the phase $f(a,b)$ depends on the
 values $c$ and $d$.)

 Let us temporarily drop the invertibility requirement for $T$. Applying the
 left hand side of Eq.~(\ref{E:XZT}) to the zero vector, we have $X_a Z_b T 0 =
 \omega_p^{f(a,b)} T X_{a\alpha + b\beta} Z_{a\delta + b\gamma} 0 =
 \omega_p^{f(a,b)} T 0$ for all $a,b\in GF(N)$. Thus, $T 0 = 0$ and hence the
 linear operator $T$ is well-defined (although it may not be invertible).

 In contrast, an invertible $T$ satisfying Eq.~(\ref{E:XZT}) does not exist in
 general. To see this, we use Eqs.~(\ref{E:Algebra_X_X})--(\ref{E:XZT}) to
 manipulate the expression $X_{a+c} Z_{b+d} T$. On one hand, $X_{a+c} Z_{b+d} T
 = \omega_p^{f(a+c,b+d)} T X_{(a+c)\alpha+(b+d)\beta} Z_{(a+c)\delta+(b+d)
 \gamma}$. On the other hand, $X_{a+c} Z_{b+d} T$ $=\omega_p^{-\textrm{Tr}(b
 c)} X_a Z_b X_c Z_d T$ $= \omega_p^{f(c,d)-\textrm{Tr}(b c)} X_a Z_b T X_{c
 \alpha+d\beta} Z_{c\delta+d\gamma}$ $=\omega_p^{f(a,b)+f(c,d)}$
 $\omega_p^{\textrm{Tr}([a\delta+b\gamma] [c\alpha+d\beta]-b c)}$ $T X_{(a+c)
 \alpha+(b+d)\beta} Z_{(a+c)\delta+(b+d)\gamma}$. Thus, the above two ways of
 expressing $X_{a+c} Z_{b+d} T$ agree for all $a,b,c,d\in GF(N)$ is a necessary
 condition for the existence of $T^{-1}$; otherwise $T$ is not injective as it
 maps a non-zero vector to the zero vector.

 It is tedious but straight-forward to check that the phase factor given in
 Eq.~(\ref{E:phase_factor}), together with the three phase conventions
 (\ref{E:phase_convention_pgt2})--(\ref{E:phase_convention_p2b}) below
 satisfy the necessary condition for the existence of $T^{-1}$ stated in the
 above paragraph. More importantly, we prove in Theorem~\ref{Thrm:Explicit_TM}
 that the phase factor $f_M(a,b)$ defined in this way makes $T(M)$ invertible
 for all $M\in SL(2,N)$. We begin by writing down this particular phase factor
 $f_M(a,b)$ below:
\begin{eqnarray}
 f_M(a,b) & = & \frac{1}{2} \textrm{Tr} ([ a^2 \alpha\delta + b^2 \beta\gamma])
 + \textrm{Tr} (a b \beta\delta) + \nonumber \\
 & & ~~\Delta_{p2} \textrm{Tr} (\sum_{i>j} g_i g_j [a_i a_j \alpha\delta + b_i
 b_j \beta\gamma ]) \label{E:phase_factor}
\end{eqnarray}
 for all $a,b\in GF(N)$. Note that in Eq.~(\ref{E:phase_factor}), $a =
 \sum_{i=1}^n a_i g_i$ and $b = \sum_{i=1}^n b_i g_i$ where $\{ g_1, g_2,
 \ldots , g_n \}$ is a fixed basis of $GF(N)$ over the field $GF(p)$ and $a_i,
 b_i \in GF(p)$. Moreover, $\Delta_{p2} = 1$ if $p=2$ and $\Delta_{p2} = 0$ if
 $p\neq 2$ in the above equation is the Kronecker delta.

 The phase conventions are chosen as follows. When $p>2$ and hence $N$ is odd,
 $2$ is invertible in $GF(N)$. Consequently, the phase $\omega_p^{f_M(a,b)}$
 may be chosen from $p$th roots of unity. Following this convention requires
\begin{equation}
 f_M(a,b) \in \mathbb{Z}/p{\mathbb Z} \textrm{~for any~} a,b\in GF(N)
 \textrm{~if~} 2\nmid N . \label{E:phase_convention_pgt2}
\end{equation}
 In contrast, when $p=2$ and hence $N$ is even, $2$ is not invertible in
 $GF(N)$. Consequently, $f_M(a,b)$ may be integral or half-integral; and
 $\omega_p^{f_M(a,b)} \in \{ \pm 1,\pm i \}$. In this case, we use the
 convention
\begin{equation}
 \omega_2^{\textrm{Tr} (g_j^2 a_j^2 \alpha\delta)/2} = \left\{
 \begin{array}{ll} 1 & \textrm{if~} \textrm{Tr} (g_j^2 a_j^2 \alpha\delta) = 0,
 \\ i & \textrm{if~} \textrm{Tr} (g_j^2 a_j^2 \alpha\delta) = 1, \end{array}
 \right. \label{E:phase_convention_p2a}
\end{equation}
 and
\begin{equation}
 \omega_2^{\textrm{Tr} (g_j^2 b_j^2 \beta\gamma)/2} = \left\{ \begin{array}{ll}
 1 & \textrm{if~} \textrm{Tr} (g_j^2 b_j^2 \beta\gamma) = 0, \\ i &
 \textrm{if~} \textrm{Tr} (g_j^2 b_j^2 \beta\gamma) = 1, \end{array} \right.
 \label{E:phase_convention_p2b}
\end{equation}
 for all $a_j,b_j\in GF(p)$, where $j=1,2,\ldots ,n$.

 We explain why the last term in Eq.~(\ref{E:phase_factor}) is required. Recall
 that the identity $\textrm{Tr} (a_i^2 + a_j^2)/2 + \textrm{Tr} (a_i a_j) =
 \textrm{Tr} ([a_i + a_j]^2)/2$ holds only for $p>2$. In contrast, $\textrm{Tr}
 (a_i^2 + a_j^2) = \textrm{Tr} ([a_i + a_j]^2)$ for $p=2$. So, we cannot use
 the first identity to absorb the last term in Eq.~(\ref{E:phase_factor}) into
 the first term when $p=2$.

\par\medskip
\begin{Lem}
 Suppose $T(M)$ is a non-zero linear operator obeying
 Eqs.~(\ref{E:phase_factor})--(\ref{E:phase_convention_p2b}) as well as the
 equation $X_a Z_b \,T(M) = \omega_p^{f_M(a,b)} T(M) \,X_{a\alpha+b\beta} Z_{a
 \delta+b\gamma}$ for all $a,b\in GF(N)$. Then $T(M)$ is invertible. Besides,
 $T(M)$ is unitary after a proper scaling. Specifically, $T(M)$ is unitary if
 and only if its operator norm satisfies $\| T(M) \| = 1$.
 \label{Lem:Unitary_Condition}
\end{Lem}
\begin{proof}
 Clearly, $T$ also satisfies the equation $T^\dagger Z_{-b} X_{-a} =
 \omega_p^{-f(a,b)} Z_{-a\delta -b\gamma} X_{-a\alpha -b\beta} T^\dagger$. From
 Eqs.~(\ref{E:phase_factor})--(\ref{E:phase_convention_p2b}), we know that $X_a
 Z_b T T^\dagger =$ $\omega_p^{f(a,b)} T X_{a\alpha + b\beta} Z_{a\delta + b
 \gamma} T^\dagger =$ $\omega_p^{f(a,b) - \textrm{Tr} ([a\alpha + b\beta ][a
 \delta + b\gamma])} T Z_{-a\delta-b\gamma}^\dagger X_{-a\alpha-b\beta}^\dagger
 T^\dagger =$ $\omega_p^{f(a,b)+f(-a,-b) - \textrm{Tr} (a^2 \alpha\delta + b^2
 \beta \gamma) - 2 \textrm{Tr} (a b \beta\delta)} T T^\dagger X_a Z_b =$ $T
 T^\dagger X_a Z_b$ for all $a,b\in GF(N)$. By the same argument, $X_a Z_b
 T^\dagger T = T^\dagger T X_a Z_b$ for all $a,b\in GF(N)$. Thus $T T^\dagger$
 and $T^\dagger T$ are non-zero operators belonging to the centralizer of $\{
 \sum_{a,b} \Lambda_{ab} X_a Z_b : \Lambda_{ab} \in {\mathbb C} \}$. In other
 words, $T T^\dagger$ and $T^\dagger T$ are non-zero constant multiples of the
 identity operator. Hence, $T$ is invertible. Obviously, the invertible
 operator $T$ is unitary if and only if $\| T \| = 1$.
\end{proof}

\par\medskip
\begin{Thrm}
 Let $\{ g_1,g_2, \ldots ,g_n \}$ be a fixed basis of $GF(N)$ over $GF(p)$. For
 any $M = \left[ \begin{array}{cc} \alpha & \beta \\ \delta & \gamma
 \end{array} \right] \in SL(2,N)$, the unitary operator $T(M)$ satisfying 
 Eqs.~(\ref{E:XZT})--(\ref{E:phase_convention_p2b}) exists. A possible choice
 of $T(M)$ is
 \begin{eqnarray}
  T(M) & \!\!\!= & \!\!\!\!\!\frac{e^{i\theta}}{N^{\dim
   (\textrm{\scriptsize colspan} (M-I))/2}} \!\!\sum_{[a~b]\in
   \textrm{\scriptsize colspan}(M-I)}
   \!\!\!\!\!\!\omega_p^{\textrm{Tr}(\varphi_M (a,b))} \times \nonumber \\
  & & ~~\omega_p^{\frac{1}{2} \textrm{Tr} (\varphi'_M (a,b))} X_a Z_{b}
  \label{E:Explicit_TM}
 \end{eqnarray}
 for some $\theta\in {\mathbb R}$, with $\textrm{colspan} (M-I)$ being the span
 of the columns of $(M-I)$. In the above equation, the functions $\varphi_M,
 \varphi'_M : GF(N)\times GF(N) \longrightarrow GF(N)$ are given by
 \begin{eqnarray}
  & & \varphi_M (a,b) \nonumber \\
  & = & b [\alpha \tilde{a}(a,b) + \beta \tilde{b}(a,b)] - a \tilde{b}(a,b) -
   \alpha\delta\tilde{a}(a,b)^2 \nonumber \\
  & & ~~-\alpha (\gamma-1) \tilde{a}(a,b) \tilde{b}(a,b) - \beta (\gamma-1)
   \tilde{b}(a,b)^2 \nonumber \\
  & & ~~+\Delta_{p2} \sum_{i>j} g_i g_j [ \alpha\delta \tilde{a}_i (a,b)
   \tilde{a}_j (a,b) \nonumber \\
  & & ~~+ \beta\gamma \tilde{b}_i (a,b) \tilde{b}_j (a,b) ]
  \label{E:Explicit_varphi1}
 \end{eqnarray}
 and
 \begin{equation}
  \varphi'_M (a,b) = \alpha\delta \tilde{a}(a,b)^2 + \beta\gamma \tilde{b}(a,
  b)^2 \label{E:Explicit_varphi2}
 \end{equation}
 respectively. In Eqs.~(\ref{E:Explicit_varphi1})
 and~(\ref{E:Explicit_varphi2}), $\tilde{a}(a,b), \tilde{b}(a,b) \in GF(N)$ and
 $\tilde{a}_i (a,b), \tilde{b}_i (a,b) \in GF(p)$ are the solutions of the
 system of equations
 \begin{equation}
  \sum_{i=1}^n g_i \tilde{a}_i (a,b) = \tilde{a} (a,b) , \label{E:tilde_a}
 \end{equation}
 \begin{equation}
  \sum_{i=1}^n g_i \tilde{b}_i (a,b) = \tilde{b} (a,b) \label{E:tilde_b}
 \end{equation}
 and
 \begin{equation}
  \left[ \begin{array}{cc} \alpha - 1 & \beta \\ \delta & \gamma - 1
  \end{array} \right] \left[ \begin{array}{c} \tilde{a}(a,b) \\ \tilde{b}(a,b)
  \end{array} \right] = \left[ \begin{array}{c} a \\ b \end{array} \right] .
  \label{E:solution_tilde_ab}
 \end{equation}
 \label{Thrm:Explicit_TM}
\end{Thrm}
\begin{proof}
 We show the existence of $T$ by explicitly constructing it. We write $T =
 \sum_{i,j\in GF(N)} \Lambda_{ij} X_i Z_j$ for some $\Lambda_{ij}\in
 {\mathbb C}$.  Substituting this $T$ into Eq.~(\ref{E:XZT}) and equating the
 coefficient of $X_a Z_b$, we obtain
 \begin{eqnarray}
  \Lambda_{ab} & = & \omega_p^{f(i,j) + \textrm{Tr} ([i\alpha + j\beta] \{b - i
  \delta - j [ \gamma -1] \} - a j)} \times \nonumber \\
  & & ~~\Lambda_{a-i(\alpha-1)-j\beta,b-i\delta-j(\gamma-1)}
  \label{E:T_relation}
 \end{eqnarray}
 for all $a,b,i,j\in GF(N)$. Using
 Eqs.~(\ref{E:phase_factor})--(\ref{E:phase_convention_p2b}), it is tedious but
 straight-forward to check that Eq.~(\ref{E:T_relation}) consists of $N^2$, $N
 (N-1)$ and $(N^2-1)$ independent equations when $(M-I)$ is of rank $0$, $1$
 and $2$ respectively.

 In what follows, we only consider the case $\det (M-I) \neq 0$. The other
 cases can be proven in a similar manner. Since $(M-I)$ is invertible, $\dim (
 \textrm{colspan} (M-I)) = 2$. Besides, the solution of $\tilde{a}(a,b),
 \tilde{b}(a,b) \in GF(N)$ and $\tilde{a}_i(a,b), \tilde{b}_i(a,b) \in GF(p)$
 in the system of Eqs.~(\ref{E:tilde_a})--(\ref{E:solution_tilde_ab}) exists
 and is unique for any given $a,b\in GF(N)$. Hence, by choosing these
 $\tilde{a}(a,b), \tilde{b}(a,b), \tilde{a}_i(a,b), \tilde{b}_i(a,b)$, we may
 use the $(N^2-1)$ independent equations taken from Eq.~(\ref{E:T_relation}) to
 relate every $\Lambda_{ab}$ to $\Lambda_{00}$ for all $(a,b)\neq (0,0)$. In
 this way, we conclude that every $\Lambda_{ab}$ is proportional to
 $\Lambda_{00}$. Besides, all $|\Lambda_{ab}|$'s are equal. Consequently, from
 Lemma~\ref{Lem:Unitary_Condition}, the unitarity of $T(M)$ implies that
 $|\Lambda_{00}| = 1/N$. Substituting $\tilde{a}(a,b), \tilde{b}(a,b)$ into
 Eqs.~(\ref{E:XZT})--(\ref{E:phase_convention_p2b}) and~(\ref{E:T_relation}),
 we arrive at Eqs.~(\ref{E:Explicit_TM})--(\ref{E:Explicit_varphi2}).
\end{proof}

\par\medskip
 For the purpose of illustration, the unitary operators $T(M)$'s for a few
 $M$'s computed by
 Eqs.~(\ref{E:phase_convention_pgt2})--(\ref{E:Explicit_varphi2}) are listed in
 Table~\ref{T:T}. Incidentally, the unitary operator $T(M)$ listed in
 Table~\ref{T:T} for $N=2$ is, up to a global phase, the same as the one used
 by Lo in his security proof of the six-state scheme in
 Ref.~\cite{sixstateproof}. Furthermore, it is shown in
 Theorem~\ref{Thrm:Subgroup} in the Appendix that the first three operators
 listed in Table~\ref{T:T} are of great importance in the construction of QKD
 schemes for $N=2,3$.

\begin{table}[t]
\begin{center}
\begin{tabular}{|c|c|l|}
 \hline
 $N$ & $M$ & $T(M)$ \\ \hline
 2 & $\left[ \begin{array}{cc} 0 & 1 \\ 1 & 1 \end{array} \right]$ & 
  $\displaystyle \frac{1}{2} \left( I + i X_1 + i Z_1 + X_1 Z_1 \right)$ \\
 \hline
 3 & $\left[ \begin{array}{cc} 1 & 1 \\ 1 & 2 \end{array} \right]$ &
  $\displaystyle \frac{1}{3} \sum_{a,b=0}^2 \omega_3^{\Delta_{b0}-\Delta_{a0}}
  X_a Z_b$ \\ \hline
 3 & $\left[ \begin{array}{cc} 1 & 2 \\ 2 & 2 \end{array} \right]$ &
  $\displaystyle \frac{1}{3} \sum_{a,b=0}^2 \omega_3^{\Delta_{a0}-\Delta_{b0}}
  X_a Z_b$ \\ \hline
 4 & $\left[ \begin{array}{cc} 0 & 1 \\ 1 & \omega \end{array} \right]$ &
  $\displaystyle \frac{1}{4} \!\!\sum_{a,b\in GF(4)} \!\!\!(-1)^{\textrm{Tr}
  (\omega [a+b])/2 + \textrm{Tr} (\omega^2 [a+b + \tilde{b}_1 \tilde{b}_2])}
  X_a Z_b$ \\
 \hline
\end{tabular}
\end{center}
\caption{The operator $T$ for a few $M$'s in the case of $N=2$, $3$ and $4$.
 Note that $\omega\in GF(4)$ in the last row of the table satisfies $\omega^2 +
 \omega + 1 = 0$.}
\label{T:T}
\end{table}

 The unitary operator $T(M)$ stated in Theorem~\ref{Thrm:Explicit_TM} depends
 on the matrix $M\in SL(2,N)$. So we may regard $T$ as a map from $SL(2,N)$ to
 $U(N)$. Let $M_i = \left[ \begin{array}{cc} \alpha_i & \beta_i \\ \delta_i &
 \gamma_i \end{array} \right] \in SL(2,N)$ for $i=1,2$. Suppose further that
 $N$ is odd. From Eq.~(\ref{E:phase_factor}), it follows that $f_{M_1 M_2}(a,b)
 = f_{M_1} (a\alpha_2 + b\beta_2,a\delta_2 + b\gamma_2) + f_{M_2}(a,b)$ for all
 $a,b\in GF(N)$. In other words, $T(M_1 M_2) = T(M_2) T(M_1)$. Hence the map $T
 : SL(2,N) \longrightarrow U(N)$ defines a faithful transposed representation
 of $SL(2,N)$ for all odd $N$. As $SL(2,N)$ is generated by two elements for
 any prime power $N$ \cite{sl2q_generators}, Alice and Bob may apply any $T(M)$
 if they can apply the two specific unitary operators corresponding to the
 generators of $SL(2,N)$. In contrast, when $N$ is even, $T$ is not a group
 representation of $SL(2,N)$. Fortunately, readers will find out in
 Section~\ref{Sec:EntScheme} that the security of all the QKD schemes reported
 in this paper do not depend on the phase $f_M(a,b)$. Therefore, in practice,
 Alice and Bob may replace $T(M_1 M_2 \cdots M_k)$ used in the QKD schemes
 reported in this paper by $T(M_k) T(M_{k-1}) \cdots T(M_1)$ in which $M_i$'s
 are chosen from the two generators of $SL(2,N)$. (Note that the unitary
 operator defined in this way may depend on the decomposition of a matrix in
 $SL(2,N)$ into factors of $M_i$'s. However, the unitary operator defined by
 any such decomposition will work equally well.)

\subsection{An Entanglement-Based Quantum Key Distribution Scheme
 \label{Subsec:EntScheme}}
\par\medskip
\begin{enumerate}[\underline{EB QKD Scheme~A}]
 \item Let the Hilbert space dimension $N$ of each quantum particle involved in
  this scheme be a prime power. Alice prepares $L\gg 1$ quantum particle pairs
  in the state $\sum_{i\in GF(N)} |ii\rangle/\sqrt{N}$. She randomly and
  independently applies a unitary transformation $T(M) \in T[SL(2,N)]$ to the
  second particle in each pair. She keeps the first particle and sends the
  second in each pair to Bob. Bob acknowledges the receipt of these particles
  and then applies a randomly and independently picked $T(M')^{-1}$ to each
  received particle. Now, Alice and Bob publicly reveal their unitary
  transformations applied to each particle. A shared pair is then kept and is
  said to be in the set $S_M$ if Alice and Bob have applied $T(M)$ and
  $T(M)^{-1}$ to the second particle of the shared pair respectively. Thus in
  the absence of noise and Eve, each pair of shared particles kept by Alice and
  Bob should be in the state $\sum_{i\in GF(N)} |ii\rangle/\sqrt{N}$.
  \label{Ent:Prepare}
 \item Alice and Bob estimate the channel error rate by sacrificing a few
  particle pairs. Specifically, they randomly pick $\mbox{O} ([N+1]^2 \log \{ N
  [N^2 -1]/\epsilon \} / \delta^2 N^2)$ pairs from each of the $N(N^2-1)$ sets
  $S_M$ and measure each particle of the pair in the $\{ |i\rangle : i\in GF(N)
  \}$ basis, namely the standard basis. They publicly announce and compare
  their measurement results. In this way, they know the estimated channel error
  rate to within $\delta$ with probability at least $(1-\epsilon)$. (A detailed
  proof of this claim can be found in Ref.~\cite{biasedbb84}. A brief outline
  of the proof will also be given in Subsection~\ref{Subsec:Sampling} for handy
  reference.) If the channel error rate is too high, they abort the scheme and
  start all over again.
  \label{Ent:QCTest}
  \item Alice and Bob perform the following privacy amplification procedure.
   (It will be shown in Section~\ref{Sec:SecEntScheme} that
   step~\ref{Ent:PriAmp_SpinCorr} below reduces errors of the form $X_a Z_b$
   with $a\neq 0$ at the expense of increasing errors of the form $Z_c$ with $c
   \neq 0$. In contrast, step~\ref{Ent:PriAmp_PhaseCorr} below reduces errors
   of the form $X_a Z_b$ with $b\neq 0$ at the expense of increasing errors of
   the form $X_c$ with $c\neq 0$. Applying steps~\ref{Ent:PriAmp_SpinCorr}
   and~\ref{Ent:PriAmp_PhaseCorr} in turn is an effective way to reduce all
   kinds of quantum errors provided that the error rate is not too high.)
   \begin{enumerate}
    \item Alice and Bob apply the entanglement purification procedure by
     two-way classical communication (LOCC2 EP) similar to the one reported in
     Refs.~\cite{bdsw,genpur}. Specifically, Alice and Bob randomly group their
     remaining quantum particles in tetrads where each tetrad consists of two
     pairs shared by Alice and Bob in Step~\ref{Ent:Prepare}. Alice randomly
     picks one of the two particles in her share of each tetrad as the control
     register and the other as the target. She applies the following unitary
     operation to the control and target registers:
     \begin{equation}
      |i\rangle_\textrm{control} \otimes |j\rangle_\textrm{target} \longmapsto
      |i\rangle_\textrm{control} \otimes |j-i\rangle_\textrm{target} ,
      \label{E:GenCNOT}
     \end{equation}
     where the subtraction is performed in the finite field $GF(N)$. Bob
     applies the same unitary transformation to his corresponding share of
     particles in the tetrad. Then, they publicly announce the measurement
     results of their target registers in the standard basis. They keep their
     control registers only when the measurement results of their corresponding
     target registers agree. They repeat the above LOCC2 EP procedure until
     there is an integer $r>0$ such that a single application of
     step~\ref{Ent:PriAmp}b will bring the signal quantum error rate of the
     resultant particles down to less than $\epsilon_I / \ell^2$ for an
     arbitrary but fixed security parameter $\epsilon_I > 0$, where $r \ell$ is
     the number of remaining pairs they share currently. They abort the scheme
     either when $r$ is greater than the number of remaining quantum pairs they
     possess or when they have used up all their quantum particles in this
     procedure. \label{Ent:PriAmp_SpinCorr}
    \item They apply the majority vote phase error correction (PEC) procedure
     introduced by Gottesman and Lo \cite{qkd2waylocc}. Specifically, Alice and
     Bob randomly divide the resultant particles into sets each containing $r$
     pairs of particles shared by Alice and Bob. Alice and Bob jointly apply
     the $[r,1,r]_N$ phase error correction procedure to their corresponding
     shares of $r$ particles in each set and retain their phase error corrected
     quantum particles. At this point, Alice and Bob should share $\ell$ almost
     perfect pairs $\sum_{i\in GF(N)} |ii\rangle/\sqrt{N}$ with fidelity at
     least $(1-\epsilon_I /\ell)$. By measuring their shared pairs in the
     standard basis, Alice and Bob obtain their common key. More importantly,
     Eve's information on this common key is less than the security parameter
     $\epsilon_I$. (Proof of this claim can be found in
     Theorem~\ref{Thrm:Uncond_Sec_A} in Subsection~\ref{Subsec:PA} below.)
     \label{Ent:PriAmp_PhaseCorr}
   \end{enumerate}
   \label{Ent:PriAmp}
\end{enumerate}

\par\medskip
 One may simplify Scheme~A by picking $T(M)$'s from $T[H]$, where $H$ is a
 proper subgroup of $SL(2,N)$ whose number of elements divides $(N^2-1)$.
 Theorem~\ref{Thrm:Subgroup} in the Appendix tells us that the subgroup $H$
 exists if and only if $N=2,3,5,7,11$ and $|H| = N^2-1$. From now on, we use
 the symbol $G$ to denote either the entire group $SL(2,N)$ or the order
 $(N^2-1)$ subgroup $H$ of $SL(2,N)$.

 In the case $N=2$ and $G$ equals the cyclic group of three elements, Scheme~A
 is a variation of the six-state scheme introduced by Chau in
 Ref.~\cite{sixstateexact}. The key difference is that, unlike the former one,
 the present scheme does not make use of Calderbank-Shor-Steane quantum code
 after PEC.

 Lemma~\ref{Lem:Depolarize} in Subsection~\ref{Subsec:PA} shows that all Pauli
 errors in the quantum signal right after step~\ref{Ent:Prepare} in Scheme~A
 are depolarized. Furthermore, Theorem~\ref{Thrm:Subgroup} in the Appendix
 shows that the same conclusion applies when Alice and Bob pick $M$ from a
 subgroup $H$ of $SL(2,N)$ of order $(N^2-1)$.

\section{Cryptanalysis Of The Entanglement-Based Quantum Key Distribution
 Scheme \label{Sec:SecEntScheme}}
 In this section, we present a detailed unconditional security proof of
 Scheme~A in the limit of a large number of quantum particles $L$ transmitted.
 We also investigate the maximum error tolerance rate of Scheme~A against the
 most general type of eavesdropping attack allowed by the laws of quantum
 physics. With suitable modifications, the security proof reported here can be
 extended to the case of a small finite $L$. Nevertheless, working in the limit
 of large $L$ makes the asymptotic error tolerance rate analysis easier.

 The remainder of this section is organized as follows. In
 Subsection~\ref{Subsec:Error_Rates}, we define various error rate measures and
 discuss how to fairly compare error tolerance capabilities between different
 QKD schemes. Then in Subsection~\ref{Subsec:Sampling}, we briefly explain why
 a reliable upper bound of the channel error can be obtained by randomly
 testing only a small subset of quantum particles in step~\ref{Ent:QCTest} of
 Scheme~A. Finally in Subsection~\ref{Subsec:PA}, we prove the security of the
 privacy amplification procedure in step~\ref{Ent:PriAmp} of Scheme~A and
 analyze its error tolerance rate. This will complete the proof of
 unconditional security for EB Scheme~A.

\subsection{Fair Comparison Of Error Tolerance Capability And Various Measures
 Of Error Rates \label{Subsec:Error_Rates}}

\par\medskip
\begin{Def}
 Recall that Alice prepares $L$ particle pairs each in the state $\sum_{i\in
 GF(N)} |ii\rangle/\sqrt{N}$ and randomly applies $T(M)\in T[G]$ to the second
 particle in each pair. We denote the resultant (pure) state of the pairs by
 $\bigotimes_{j=1}^L |\phi_j\rangle$. Then, she sends one particle in each pair
 through an insecure quantum channel to Bob; and upon receipt, Bob randomly
 applies $T(M')^{-1}$ to his share of the pair. The {\bf channel quantum error
 rate} in this situation is defined as the marginal error rate of the
 measurement results if Alice and Bob were to make an hypothetical measurement
 on the $j$th shared quantum particle pair in the basis $\{ I \otimes X_a Z_b
 |\phi_j\rangle : a,b\in GF(N) \}$ for all $j$. In other words, the channel
 quantum error rate equals $1/L$ times the expectation value of the cardinality
 of the set $\{ j : \textrm{hypothetical measurement of the $j$th pair equals }
 I \otimes X_a Z_b |\phi_j\rangle \textrm{ with } (a,b)\neq (0,0) \}$. The
 {\bf channel standard basis measurement error rate} is defined as $1/L$ times
 the expectation value of the cardinality of the set $\{ j :
 \textrm{hypothetical measurement of the $j$th pair equals } I \otimes X_a Z_b
 |\phi_j\rangle \textrm{ with } a\neq 0 \}$. The next two definitions concern
 only those quantum particle pairs retained by Alice and Bob in $\bigcup_{M\in
 G} S_M$. (That is, those that Alice and Bob have applied $T(M)$ and
 $T(M)^{-1}$ to the second particle of the shared pair for some $M\in G$
 respectively.) In the absence of noise and Eve, all such particle pairs should
 be in the state $\sum_{i\in GF(N)} |ii\rangle/\sqrt{N}$. The {\bf signal
 quantum error rate} (or quantum error rate (QER) for short) in this situation
 is defined as the expectation value of the proportion of particle pairs in
 $\bigcup_M S_M$ whose measurement result in the basis $\{ \sum_{i\in GF(N)}
 |i\rangle \otimes X_a Z_b |i\rangle/\sqrt{N}: a,b\in GF(N) \}$ equals
 $\sum_{i\in GF(N)} |i\rangle \otimes X_a Z_b |i\rangle/\sqrt{N}$ for some
 $(a,b) \neq (0,0)$. The {\bf signal standard basis measurement error rate} (or
 standard basis measurement error rate (SBMER) for short) is defined as the
 expectation value of the proportion of particle pairs in $\bigcup_M S_M$ whose
 measurement result in the basis $\{ \sum_{i\in GF(N)} |i\rangle \otimes X_a
 Z_b |i\rangle/\sqrt{N} : a,b\in GF(N) \}$ equals $\sum_{i\in GF(N)} |i\rangle
 \otimes X_a Z_b |i\rangle/\sqrt{N}$ for some $a\neq 0$. In other words, SBMER
 measures the apparent error rate of the signal when Alice and Bob measure
 their respective shares of particles in the standard basis. In the special
 case of $N = 2^n$, any standard basis measurement result can be bijectively
 mapped to an $n$-bit string. Thus, it makes sense to define the {\bf signal
 bit error rate} (or bit error rate (BER) for short) as the marginal error rate
 of the $n$-bit string resulting from a standard basis measurement of the
 signal at the end of the signal preparation and transmission stage.
 \label{Def:Error_Rates}
\end{Def}

\par\medskip
 Three important remarks are in place. First, SBMERs and BERs of QKD schemes
 using quantum particles of different dimensions as information carriers should
 \emph{never} be compared directly. This is because the quantum communication
 channels used are different. In addition, the same eavesdropping strategy may
 lead to different error rates
 \cite{alph,highdim,dlevel,highdimmore,evethreestate}. It appears that the only
 sensible situation in which it is meaningful compare the error tolerance
 capabilities of two QKD schemes is when the schemes are using the same quantum
 communication channel and are subjected to the same eavesdropping attack.
 Specifically, let Alice reversibly map every $p^n$-dimensional quantum state
 used in Scheme~A into $n$ possibly entangled $p$-dimensional quantum particles
 and send them through an insecure $p$-dimensional quantum particle
 communication channel to Bob. Moreover, since we assume that Alice and Bob do
 not have quantum storage capability, it is reasonable to require that Alice
 prepares and sends packets of $n$ possibly entangled $p$-dimensional quantum
 particles one after another. In this way, Scheme~A becomes an
 entangled-particle-based QKD scheme. More importantly, Eve may apply the same
 eavesdropping attack on the insecure $p$-dimensional quantum particle channel
 used by Alice and Bob irrespective of the value $n$. Thus, it is fair to
 compare the error tolerance capability between two entangled-particle-based
 QKD schemes derived from Scheme~A using $p^n$- and $p^{n'}$-dimensional
 particles respectively against any eavesdropping attack on the $p$-dimensional
 quantum particle channel.

 Second, the BER defined above for $N=2^n$ with $n>1$ depends on the bijection
 used. Fortunately, in Subsection~\ref{Subsec:PA}, readers will find that the
 BER for the QKD scheme reported in this paper is independent of this
 bijection.

 Third, Lemma~\ref{Lem:Depolarize} in Subsection~\ref{Subsec:PA} and
 Theorem~\ref{Thrm:Subgroup} in the Appendix show that Pauli errors that
 occurred in a collection of $N$-dimensional quantum registers are depolarized
 if we conjugate each register by a randomly and independently picked $T(M) \in
 T[G]$. Furthermore, the channel quantum error rate is equal to the QER of the
 signal. Roughly speaking, QER refers to the rate of any quantum error (phase
 shift and/or spin flip) occurring in the pair $\sum_{i\in GF(N)} |ii\rangle /
 \sqrt{N}$ shared by Alice and Bob. In contrast, the depolarization of Pauli
 errors implies that the channel standard basis measurement error rate does not
 equal the SBMER in general.

\subsection{Reliability Of The Error Rate Estimation \label{Subsec:Sampling}}
 In Scheme~A, Alice and Bob keep only those particle pairs that are believed to
 be in the state $\sum_{i\in GF(N)} |ii\rangle / \sqrt{N}$ at the end of
 step~\ref{Ent:Prepare}. Then, they measure some of them in the standard basis
 in the signal quality control test in step~\ref{Ent:QCTest}. More importantly,
 since all the LOCC2 EP and PEC privacy amplification procedures in
 step~\ref{Ent:PriAmp} map standard basis to standard basis, we can imagine
 that the final standard basis measurements of their shared secret key were
 performed right at the beginning of step~\ref{Ent:PriAmp}. In this way, any
 quantum eavesdropping strategy used by Eve is reduced to a classical
 probabilistic cheating strategy. In other words, for any quantum eavesdropping
 strategy, one can always find an equivalent Pauli attack that has the same
 probability of passing the signal quality control test in
 step~\ref{Ent:QCTest} and gives the same density matrix of the shared quantum
 particles just before the final standard basis measurement in
 step~\ref{Ent:PriAmp}. Therefore, we need only to consider Pauli attack in the
 subsequent analysis \cite{lochauqkdsec}. 

 Recall that in step~\ref{Ent:QCTest}, Alice and Bob do not care about the
 measurement result of an individual quantum register; they only care about the
 difference between the measurement outcome of Alice and the corresponding
 outcome of Bob. In other words, they apply the projection operator
\begin{equation}
 P_a = \sum_{i\in GF(N)} |i,i+a\rangle \,\langle i,i+a| \label{E:P_form_ind}
\end{equation}
 to each of the randomly selected quantum registers in the set $\bigcup_{M\in
 G} S_M$. The projection operator $P_a$ can be rewritten in a form involving
 Bell-like states as follows. Define $|\Phi_{ab}\rangle$ to be the Bell-like
 state $\sum_{i\in GF(N)} |i\rangle \otimes X_a Z_b |i\rangle / \sqrt{N} \equiv
 \sum_{i\in GF(N)} \omega_p^{\textrm{Tr} (i b)} |i,i+a\rangle / \sqrt{N}$.
 Then, $P_a$ can be rewritten as
\begin{equation}
 P_a = \sum_{b\in GF(N)} |\Phi_{ab}\rangle \,\langle\Phi_{ab}| .
 \label{E:P_form_Bell}
\end{equation}
 Since every particle pair in $S_M$ is subjected to $T(M)$ and $T(M)^{-1}$
 before and after passing through the insecure channel respectively, $P_a$ is a
 measure of whether an error of the form $T(M) X_a Z_b T(M)^{-1}$ for some $b
 \in GF(N)$ has occurred in this pair. Recall that $M\in G$ is randomly and
 independently chosen for each pair. Moreover, such a choice is known to Eve
 after the second half of the particle pair has reached Bob. So, combined with
 Eqs.~(\ref{E:XZT}) and (\ref{E:P_form_ind})--(\ref{E:P_form_Bell}), the signal
 quality control test in step~\ref{Ent:QCTest} of Scheme~A can be regarded as
 an effective random sampling test for the fidelity of the pairs as $|\Phi_{00}
 \rangle \equiv \sum_{i\in GF(N)} |ii\rangle/\sqrt{N}$.  

 At this point, classical sampling theory can be used to estimate the quantum
 channel error and hence the eavesdropping rate of the classical probabilistic
 cheating strategy used by Eve, as well as the fidelity of the remaining pairs
 as $|\Phi_{00}\rangle$.

\par\medskip
\begin{Lem}[Adapted from Lo, Chau and Ardehali \cite{biasedbb84}]
 Suppose that immediately after step~\ref{Ent:Prepare} in Scheme~A, Alice and
 Bob share $L_M$ pairs of particles in the set $S_M$, namely, those particles
 that were conjugated by $T(M)$. Suppose further that Alice and Bob randomly
 pick $\mbox{O} (\log [ 1/ \epsilon] / \delta^2) \lesssim 0.01 L_M$ of the
 $L_M$ pairs for testing in step~\ref{Ent:QCTest}. Define the estimated channel
 standard basis measurement error rate $\hat{e}_M$ to be the portion of tested
 pairs whose measurement results obtained by Alice and Bob differ. Denote the
 channel standard basis measurement error rate for the set $S_M$ by $e_M$.
 Then, the probability that $|e_M - \hat{e}_M| > \delta$ is of the order of
 $\epsilon$ for any fixed $\delta > 0$. \label{Lem:Sampling}
\end{Lem}
\begin{proof}
 Using earlier discussions in this subsection, the problem depicted in this
 lemma is equivalent to a classical random sampling problem without
 replacement whose solution follows directly from Lemma~1 in
 Ref.~\cite{biasedbb84}.
\end{proof}

\par\medskip
 Lemma~\ref{Lem:Sampling} assures that by randomly choosing $\mbox{O} (\log [1
 / \epsilon] / \delta^2)$ out of $L_M$ pairs to test, the unbiased estimator
 $\hat{e}_M$ cannot differ significantly from the actual channel standard basis
 measurement error rate $e_M$. More importantly, the number of particle pairs
 they need to test is independent of $L_M$. Therefore, in the limit of large
 $L_M$ (and hence large $L$), randomly testing a negligibly small portion of
 quantum particle pairs is sufficient for Alice and Bob to estimate the channel
 standard basis measurement error rate in the set $S_M$ with high confidence
 \cite{biasedbb84}. In addition, the QER of the remaining untested particle
 pairs is the same as that of $\bigcup_{M\in G} S_M$ in the large $L$ limit.

\par\medskip
\begin{Thrm}
 Let $G$ denote the group $SL(2,N)$ or its order $(N^2-1)$ subgroup $H$
 reported in Theorem~\ref{Thrm:Subgroup}. Using the notation in
 Lemma~\ref{Lem:Sampling}, $(N+1) \left< \hat{e}_M \right> / N$ is a reliable
 estimator of the upper bound of the QER, where $\left< \cdot \right>$ denotes
 the mean averaged over all $M\in G$. Specifically, the probability that the
 QER exceeds $(N+1) ( \left< \hat{e}_M \right> + \delta ) / N$ is less than
 $\epsilon |G|$. \label{Thrm:Sampling}
\end{Thrm}
\begin{proof}
 Recall that Eve does not know the choice of unitary operators applied by Alice
 and Bob in step~\ref{Ent:Prepare} in Scheme~A. Consequently, by
 Lemma~\ref{Lem:Depolarize} in Subsection~\ref{Subsec:PA} or
 Theorem~\ref{Thrm:Subgroup} in the Appendix, step~\ref{Ent:Prepare} in
 Scheme~A depolarizes Pauli errors of the quantum particles. That is to say, in
 the limit of a large $L$, the $X_a Z_b$ error rate in the set $S_I$ is equal
 to that of $T(M)^{-1} X_a Z_b T(M)$ in the set $S_M$ for all $M\in G$. Among
 the $T(M)^{-1} X_a Z_b T(M) \equiv \omega_p^{f_M(a,b)} X_c Z_d$ errors
 occurring in the set $S_M$, only those with $c\neq 0$ can be recorded in
 step~\ref{Ent:QCTest}. Thus, the estimator for the QER equals $(N^2-1) \left<
 \hat{e}_M \right> / N(N-1) = (N+1) \left< \hat{e}_M \right> / N$. This theorem
 now follows directly from Lemma~\ref{Lem:Sampling}.
\end{proof}

\par\medskip
 To summarize, once the signal quality control test in step~\ref{Ent:QCTest} of
 Scheme~A is passed, Alice and Bob have high confidence (of at least $(1-
 \epsilon)$) that the QER of the remaining untested particle pairs is small
 enough for the signal privacy amplification stage in step~\ref{Ent:PriAmp} to
 handle. Moreover, the estimation given in Theorem~\ref{Thrm:Sampling} is
 independent of the phase $f_M(a,b)$ used by the unitary operator $T(M)$.

 Before drawing a close to this subsection, we would like to point out that one
 can estimate the QER in a more aggressive way. Specifically, Alice and Bob do
 not only know whether the measurement results of each tested pair are equal,
 in fact they also know the difference between their measurement results in
 each tested pair. They may exploit this extra piece of information to better
 estimate the probability of $X_a Z_b$ error in the signal for each $a,b\in
 GF(N)$. Such estimation helps them to devise tailor-made privacy amplification
 schemes that tackle the specific kind of error caused by channel noise and
 Eve. While this methodology will be useful in practical QKD, we shall not
 pursue this further here as the aim of this paper is the worst-case
 cryptanalysis in the limit of a large number of quantum particle transfers
 $L$.

\subsection{Security Of Privacy Amplification \label{Subsec:PA}}
\par\medskip
\begin{Def}
 We denote the $X_a Z_b$ error rate of the quantum particles shared by Alice
 and Bob just before step~\ref{Ent:PriAmp} in Scheme~A by $e_{a,b}$. When there
 is no possible confusion in the subscript, we shall write $e_{ab}$ instead of
 $e_{a,b}$. Similarly, we denote the $X_a Z_b$ error rate of the resultant
 quantum particles shared by them after $k$ rounds of LOCC2 EP by $e_{a,b}^{k
 \,\textrm{EP}}$ or $e_{ab}^{k\,\textrm{EP}}$. Suppose further that Alice and
 Bob perform PEC using the $[r,1,r]_N$ majority vote code after $k$ rounds of
 LOCC2 EP. We denote the resultant $X_a Z_b$ error rate by
 $e_{a,b}^\textrm{PEC}$ or $e_{ab}^\textrm{PEC}$. \label{Def:e_ab}
\end{Def}

\par\medskip
\begin{Lem}
 Let $G = SL(2,N)$. The signal quantum error suffered by quantum particle pairs
 in $\bigcup_{E\in SL(2,N)}$ can be regarded as depolarized. In other words,
 the QER satisfies
 \begin{equation}
  \sum_{i,j\in GF(N)} e_{ij} = 1 \label{E:Constraint_e_ab_trivial}
 \end{equation}
 and
 \begin{equation}
  e_{ab} = e_{a'b'} \mbox{~for~all~} (a,b), (a',b') \neq (0,0) .
  \label{E:Constraint_e_ab}
 \end{equation}
 \label{Lem:Depolarize}
\end{Lem}
\begin{proof}
 Recall that Alice and Bob randomly and independently apply $T(M)$ and
 $T(M')^{-1}$ to each transmitted quantum register. More importantly, their
 choices are unknown to Eve when the quantum particle is traveling in the
 insecure channel. Let ${\mathcal E}$ be the quantum operation that Eve applies
 to the quantum particles in the set $\bigcup_{M\in SL(2,N)} S_M$. (In other
 words, ${\mathcal E}$ is a completely positive convex-linear map acting on the
 set of density matrices describing the quantum particle pairs to which Alice
 and Bob have applied $T(M)$ and $T(M)^{-1}$ respectively for some $M\in
 SL(2,N)$. Moreover, $0 \leq \mbox{Tr} ({\mathcal E} (\rho)) \leq 1$ for any
 density matrix $\rho$.) After Alice and Bob have publicly announced their
 choices of quantum operations, every quantum particle pair in $\bigcup_M S_M$
 has an equal chance of having experienced $[\otimes_j T(M_j)^{-1}]
 {\mathcal E} [\otimes_j T(M_j)]$ where $M_j \in SL(2,N)$. Note that the index
 $j$ in the tensor product in the above expression runs over all particle pairs
 in $\bigcup_M S_M$. From the discussions in Subsection~\ref{Subsec:Sampling},
 we know that Eve's attack may be reduced to a classical probabilistic one. In
 other words, we may regard ${\mathcal E}$ as a Pauli error operator. Since
 $SL(2,N)$ is a group and the set $\{ M\in SL(2,N) : M [a~b]^t = [c~d]^t \}$
 contains $N$ elements for all $[a~b], [c~d] \neq [0~0]$, we conclude from
 Eq.~(\ref{E:XZT}) that the Pauli quantum error of the quantum particles in the
 set $\bigcup_{M\in SL(2,N)} S_M$ is depolarized. Hence,
 Eqs.~(\ref{E:Constraint_e_ab_trivial}) and~(\ref{E:Constraint_e_ab}) apply.
\end{proof}

\par\medskip
 After establishing the initial conditions for the QER, we investigate the
 effect of LOCC2 EP on the QER.

\par\medskip
\begin{Lem}
 In the limit of a large number of transmitted quantum registers, $e_{ab}^{k
 \,\textrm{EP}}$ is given by
 \begin{equation}
  e_{ab}^{k \,\textrm{EP}} = \frac{\sum_{c_0,\ldots ,c_{2^k-2}} e_{ac_0}
  e_{ac_1} \cdots e_{ac_{2^k-2}} e_{a,b-c_0-c_1-\cdots -c_{2^k-2}}}{\sum_{i\in
  GF(N)} \left( \sum_{j\in GF(N)} e_{ij} \right)^{2^k}} ~. \label{E:iter_e_ab}
 \end{equation}
 In particular, if $e_{ab}$'s are given by
 Eqs.~(\ref{E:Constraint_e_ab_trivial}) and~(\ref{E:Constraint_e_ab}), then
 \begin{equation}
  e_{00}^{k \,\textrm{EP}} = \frac{\left[ e_{00} + (N-1) e_{01} \right]^{2^k} +
  (N-1) \left( e_{00} - e_{01} \right)^{2^k}}{N \left\{ \left[ e_{00} + (N-1)
  e_{01} \right]^{2^k} + (N-1) N^{2^k} e_{01}^{2^k} \right\}} ,
  \label{E:e_00_kEP}
 \end{equation}
 \begin{equation}
  e_{0b}^{k \,\textrm{EP}} = \frac{\left[ e_{00} + (N-1) e_{01} \right]^{2^k} -
  \left( e_{00} - e_{01} \right)^{2^k}}{N \left\{ \left[ e_{00} + (N-1) e_{01}
  \right]^{2^k} + (N-1) N^{2^k} e_{01}^{2^k} \right\}} \label{E:e_01_kEP}
 \end{equation}
 for all $b\neq 0$ and
 \begin{equation}
  e_{ab}^{k \,\textrm{EP}} = \frac{N^{2^k} e_{01}^{2^k}}{N \left\{ \left[
  e_{00} + (N-1) e_{01} \right]^{2^k} + (N-1) N^{2^k} e_{01}^{2^k} \right\}}
  \label{E:e_10_kEP}
 \end{equation}
 for all $a,b\in GF(N)$ with $a\neq 0$. \label{Lem:EP}
\end{Lem}
\begin{proof}
 Suppose that Bob's control and target registers experience $X_a Z_b$ and
 $X_{a'} Z_{b'}$ errors respectively. (In contrast, those retained by Alice are
 error-free as they never passed through the insecure noisy channel.) After
 applying the unitary operation in Eq.~(\ref{E:GenCNOT}), the errors in the
 control and target registers become $X_a Z_{b+b'}$ and $X_{a'-a} Z_{b'}$
 respectively.

 Recall that the privacy amplification procedure in step~\ref{Ent:PriAmp} is
 performed irrespective of which set $S_M$ the particle belongs to. So, in the
 limit of a large number of transmitted quantum registers, the covariance
 between probabilities of picking any two distinct quantum registers tends to
 zero. Likewise, the covariance between probabilities of picking any two
 distinct pairs of quantum registers also tends to zero. Hence, in this limit,
 the expectation value of the $X_a Z_b$ error rate just after applying the
 unitary operation in Eq.~(\ref{E:GenCNOT}) can be computed by assuming that
 the error in every pair of control and target registers is independent.
 Moreover, the variance of the $X_a Z_b$ error rate tends to zero in this
 limit.

 To show that Eq.~(\ref{E:iter_e_ab}) is valid, let us recall that Alice and
 Bob keep their control registers only when the measurement results of their
 corresponding target registers agree. In other words, they keep a control
 register only when $a = a'$. Thus, once the control register in Bob's
 laboratory is kept, it will suffer an error $X_d Z_c$ where $d = a$ and $c = b
 + b'$. Therefore, in the limit of a large number of transmitted quantum
 registers, the number of quantum registers remaining after $(k+1)$ rounds of
 LOCC2 EP is proportional to $\sum_{i\in GF(N)} ( \sum_{j\in GF(N)} e_{ij}^{k
 \,\textrm{EP}} )^2$. Similarly, the number of quantum registers suffering from
 $X_a Z_b$ errors after $(k+1)$ rounds of LOCC2 EP is proportional to $\sum_{c
 \in GF(N)} e_{ac}^{k \,\textrm{EP}} e_{a,b-c}^{k \,\textrm{EP}}$. Furthermore,
 the two proportionality constants are the same. Therefore,
 \begin{equation}
  e_{ab}^{(k+1) \,\textrm{EP}} = \frac{\sum_{c \in GF(N)} e_{ac}^{k
  \,\textrm{EP}} e_{a,b-c}^{k \,\textrm{EP}}}{\sum_{i\in GF(N)} \left( \sum_{j
  \in GF(N)} e_{ij}^{k \,\textrm{EP}} \right)^2} \label{E:e1_ab}
 \end{equation}
 for all $k\in {\mathbb N}$. Eq.~(\ref{E:iter_e_ab}) can then be proven by
 mathematical induction on $k$. (It is easier to use mathematical induction to
 prove the validity of the numerator in Eq.~(\ref{E:iter_e_ab}) and then use
 Eq.~(\ref{E:Constraint_e_ab_trivial}) to determine the denominator.)

 In particular, if the initial error rates $e_{ab}$'s are given by
 Eqs.~(\ref{E:Constraint_e_ab_trivial}) and~(\ref{E:Constraint_e_ab}), then
 Eqs.~(\ref{E:e_00_kEP})--(\ref{E:e_10_kEP}) can be proven by mathematical
 induction on $k$ with the help of Eq.~(\ref{E:e1_ab}).
\end{proof}

\par\medskip
 Lemma~\ref{Lem:EP} generalizes a similar result for qubits
 \cite{qkd2waylocc,sixstateexact}. In fact, the effect of LOCC2 EP is to reduce
 errors of the form $X_a Z_b$ with $a\neq 0$ at the expense of possibly
 increasing errors of the form $Z_c$ with $c\neq 0$. We further remark that in
 the case where $L$ is finite, $e_{ab}^{k \,\textrm{EP}}$ is determined by
 solving the classical problem of randomly pairing $N^2$ kinds of balls in an
 urn containing $2r\ell$ balls. Therefore, $e_{ab}^{k \,\textrm{EP}}$ is
 related to the so-called multivariate hypergeometric distribution whose theory
 is reviewed extensively in Ref.~\cite{multidist}.

\par\medskip
\begin{Lem}
 In the limit of a large number of quantum particles transmitted from Alice to
 Bob, the $X_a Z_b$ error rate after PEC $e_{ab}^\textrm{PEC}$ using
 $[r,1,r]_N$ majority vote code satisfies
 \begin{equation}
  \sum_{a\neq 0} \sum_{b\in GF(N)} e_{ab}^\textrm{PEC} \leq r \sum_{a\neq 0}
  \sum_{b\in GF(N)} e_{ab}^{k \,\textrm{EP}} . \label{E:PEC_spin_flip}
 \end{equation}
 Moreover, if $e_{ab}$'s satisfy Eqs.~(\ref{E:Constraint_e_ab_trivial}),
 (\ref{E:Constraint_e_ab}) and $e_{00} > e_{01}$, then
 \begin{eqnarray}
  & & \sum_{a\in GF(N)} \sum_{b\neq 0} e_{ab}^\textrm{PEC} \nonumber \\
  & \leq & (N-1) \left\{ 1 - \frac{N \left( e_{00} - e_{01} \right)^{2^{k+
   1}}}{4\left[ e_{00} + (N-1) e_{01} \right]^{2^{k+1}}} \right\}^r
  \label{E:PEC_inequality}
 \end{eqnarray} 
 as $k\rightarrow\infty$. This inequality also holds if $r$ depends on $k$.
 \label{Lem:PEC}
\end{Lem}
\begin{proof}
 Recall that the parity check matrix of the $[r,1,r]_N$ majority vote code is
 \begin{equation}
  \left[ \begin{array}{ccccc}
   1 & -1 \\
   1 & & -1 \\
   \vdots & & & \ddots \\
   1 & & & & -1
  \end{array} \right] . \label{E:majority_vote_syndrome}
 \end{equation}
 Therefore, after measuring the (phase) error syndrome, the $Z_b$ error stays
 with the control register whereas the $X_a$ error propagates from the control
 as well as all target registers to the resultant control quantum register
 \cite{hd_ftqc}. Specifically, let the error in the $i$th quantum register be
 $X_{a_i} Z_{b_i}$ for $i=1,2,\ldots ,r$. Then, after measuring the error
 syndrome, the resultant error in the remaining control register equals $X_{a_1
 + \cdots + a_r} Z_{b_1}$. Consequently, after PEC, the error in the remaining
 register is $X_{a_1 + \cdots + a_r} Z_b$ where $b$ is the majority of $b_i$
 ($i=1,2,\ldots ,r$). In other words, after PEC, spin flip error rates are
 increased by at most $r$ times. Hence, Eq.~(\ref{E:PEC_spin_flip}) holds.

 By the same argument used in Lemma~\ref{Lem:EP}, in the limit of a large
 number of transferred quantum registers, the rate of any kind of phase error
 after PEC, $\sum_{a\in GF(N)} \sum_{b\neq 0} e_{ab}^\textrm{PEC}$, satisfies
 \begin{eqnarray}
  & & \sum_{a\in GF(N)} \sum_{b\neq 0} e_{ab}^\textrm{PEC} \nonumber \\
  & \leq & (N-1) \max \{ \textrm{Pr} \,( \textrm{the number of registers
   suffering} \nonumber \\
  & & ~\textrm{from error of the form } X_i Z_1 \textrm{ is greater than or}
   \nonumber \\
  & & ~\textrm{equal to those suffering from error of the form } X_i
   \nonumber \\
  & & ~\textrm{when drawn from a random sample of } r \textrm{ registers,}
   \nonumber \\
  & & ~\textrm{given a fixed } e_{00}) \} , \label{E:PEC_inequality1}
 \end{eqnarray}
 where the maximum is taken over all possible probabilities with different
 $e_{ab}$'s satisfying the constraints in
 Eqs.~(\ref{E:Constraint_e_ab_trivial}) and~(\ref{E:Constraint_e_ab}). We
 denote the sum $\sum_{a\in GF(N)} e_{ab}^{k\,\textrm{EP}}$ by $e_{Z_b}^{k
 \,\textrm{EP}}$. Then,
 \begin{eqnarray}
  & & \sum_{a\in GF(N)} \sum_{b\neq 0} e_{ab}^\textrm{PEC} \nonumber \\
  & \leq & (N-1) \max \{ \sum_{s=0}^r \binom{r}{s} (1-e_{Z_0}^{k \,\textrm{EP}}
   - e_{Z_1}^{k \,\textrm{EP}})^{r-s} \times \nonumber \\
  & & ~(e_{Z_0}^{k \,\textrm{EP}} + e_{Z_1}^{k \,\textrm{EP}})^s \,\textrm{Pr}
   ( \textrm{the number of registers suffering} \nonumber \\
  & & ~\textrm{from error of the form } X_i Z_1 \textrm{ is greater than or}
   \nonumber \\
  & & ~\textrm{equal to those suffering from error of the from } X_i \nonumber
   \\
  & & ~\textrm{when drawn from a random sample of } s \textrm{ registers,}
   \nonumber \\
  & & ~\textrm{given that these } s \textrm{ registers are suffering from}
   \textrm{ error} \nonumber \\
  & & ~\textrm{of the form } X_i Z_b \textrm{ for } b = 0,1, \textrm{ for}
   \textrm{ a fixed } e_{00}) \} \nonumber \\
  & \leq & (N-1) \max \{ \sum_{s=0}^r \binom{r}{s} (1-e_{Z_0}^{k \,\textrm{EP}}
   -e_{Z_1}^{k \,\textrm{EP}})^{r-s} \times \nonumber \\
  & & ~(e_{Z_0}^{k \,\textrm{EP}} + e_{Z_1}^{k \,\textrm{EP}})^s \exp \left[ -2
   s \left( \frac{1}{2}-\frac{e_{Z_1}^{k \,\textrm{EP}}}{e_{Z_0}^{k
   \,\textrm{EP}} + e_{Z_1}^{k \,\textrm{EP}}} \right)^2 \right] \} \nonumber
   \\
  & = & (N-1) \max \{ \left\{ 1 - (e_{Z_0}^{k \,\textrm{EP}} + e_{Z_1}^{k
   \,\textrm{EP}}) \times \right. \nonumber \\
  & & ~\left. \left[ e^{-2 [ 1/2 - e_{Z_1}^{k \,\textrm{EP}}/(e_{Z_0}^{k
   \,\textrm{EP}}+e_{Z_1}^{k \,\textrm{EP}})]^2} - 1 \right] \right\}^r \}
   \nonumber \\
  & \leq & (N-1) \max \{ \left[ 1- 2 t (e_{Z_0}^{k \,\textrm{EP}} + e_{Z_1}^{k
   \,\textrm{EP}}) \times \right. \nonumber \\
  & & ~\left. \left( \frac{1}{2} - \frac{e_{Z_1}^{k \,\textrm{EP}}}{e_{Z_0}^{k
   \,\textrm{EP}}+e_{Z_1}^{k \,\textrm{EP}}} \right)^2 \right]^r \}
   \label{E:PEC_inequality2}
 \end{eqnarray}
 where $t\rightarrow 1$ as $k\rightarrow\infty$. Note that we have used
 Eq.~(1.2.5) in Ref.~\cite{Roman} to arrive at the second inequality above.
 (Eq.~(1.2.5) is applicable because the assumption that $e_{00} > e_{01}$ leads
 to $e_{Z_0}^{k \,\textrm{EP}} > e_{Z_1}^{k \,\textrm{EP}}$ for a sufficiently
 large $k$.) It is straight-forward to check that Eq.~(\ref{E:PEC_inequality2})
 remains valid if $r$ depends on $k$.

 Since $e_{00} > e_{01}$, $(\sum_{b\in GF(N)} e_{0b})^{2^k} = [e_{00} + (N-1)
 e_{01}]^{2^k}$ is the dominant term in the common denominator of
 Eqs.~(\ref{E:e_00_kEP})--(\ref{E:e_10_kEP}) when $k$ is sufficiently large,
 Eq.~(\ref{E:PEC_inequality}) follows directly from
 Eqs.~(\ref{E:e_00_kEP})--(\ref{E:e_10_kEP}) and~(\ref{E:PEC_inequality2}).
\end{proof}

\par\medskip
 The above theorem tells us that the effect of PEC is to reduce errors of the
 form $X_a Z_b$ with $b\neq 0$ at the expense of possibly increasing errors of
 the form $X_c$ with $c\neq 0$. For this reason, powerful signal privacy
 amplification procedures can be constructed by suitably combining LOCC2 EP and
 PEC.

\par\medskip
 Now, we prove the unconditional security of Scheme~A.

\par\medskip
\begin{Thrm}
 Let $N = p^n$ be a prime power, and let $\epsilon_p$, $\epsilon_I$ and
 $\delta$ be three arbitrarily small but fixed positive numbers. Define
 \begin{equation}
  e^\textrm{QER} = \frac{(N^2-1)(2N+1-\sqrt{5})}{2N(N^2+N-1)} .
  \label{E:critical}
 \end{equation}
 The EB QKD Scheme~A involving the transfer of $N$-dimensional quantum
 particles is unconditionally secure with security parameters $(\epsilon_p,
 \epsilon_I)$ when the number of quantum register transfers $L \equiv
 L(\epsilon_p,\epsilon_I,\delta)$ is sufficiently large. Specifically, provided
 that Alice and Bob abort the scheme whenever the estimated QER in
 step~\ref{Ent:QCTest} is greater than $(e^\textrm{QER} - \delta)$, the secret
 key generated by Alice and Bob is provably secure in the $L\rightarrow\infty$
 limit. In fact, if Eve uses an eavesdropping strategy with at least
 $\epsilon_p$ chance of passing the signal quality test stage in
 step~\ref{Ent:QCTest}, the mutual information between Eve's measurement
 results after eavesdropping and the final secret key is less than
 $\epsilon_I$. In this respect, Scheme~A tolerates asymptotically up to a QER
 of $e^\textrm{QER}$. \label{Thrm:Uncond_Sec_A}
\end{Thrm}
\begin{proof}
 By picking $L \gg (N+1)^2 |G| \log (|G|/\epsilon_p)/\delta^2 N^2$ and applying
 Lemma~\ref{Lem:Sampling} and Theorem~\ref{Thrm:Sampling}, we conclude that by
 testing $\mbox{O} ([N+1]^2 \log [\,|G|/\epsilon_p ] / \delta^2 N^2)$ pairs in
 each set $S_M$, any eavesdropping strategy that causes a QER higher than
 $e^\textrm{QER}$ has less than $\epsilon_p$ chance of passing the signal
 quality test stage in step~\ref{Ent:QCTest} of Scheme~A. (Similarly, if the
 QER is less than $(e^\text{QER}-2\delta)$, it has at least $(1-\epsilon_p)$
 chance of passing step~\ref{Ent:QCTest}. As $\delta$ can be chosen to be
 arbitrarily small, the signal quality test stage in step~\ref{Ent:QCTest} of
 Scheme~A is not overly conservative.)

 Now, suppose that Alice and Bob arrive at the signal privacy amplification
 stage in step~\ref{Ent:PriAmp} of Scheme~A. Since $L\rightarrow\infty$, the
 quantum particle pairs used in the signal quality test stage in
 step~\ref{Ent:QCTest} do not affect the error rates $e_{ab}$'s of the
 remaining untested particle pairs.

 From the discussions in Subsection~\ref{Subsec:Sampling}, we only need to
 consider the case when Eve uses a classical cheating strategy. Hence, the
 initial error rates $e_{ab}$'s satisfy Eqs.~(\ref{E:Constraint_e_ab_trivial})
 and~(\ref{E:Constraint_e_ab}). After applying $k$ rounds of LOCC2 EP, Alice
 and Bob may consider picking $r$ used in the majority vote PEC to be 
 \begin{equation}
  r \approx \frac{\epsilon_I \left[ e_{00} + (N-1) e_{01} \right]^{2^k}}{2\ell
  (N-1) N^{2^k} e_{01}^{2^k}} , \label{E:r_bound}
 \end{equation}
 where $\ell$ is the number of quantum particle pairs Alice and Bob share
 immediately after the PEC procedure in step~\ref{Ent:PriAmp}b. Provided that
 $e_{00} > e_{01}$, in the $k\rightarrow \infty$ limit, $r\rightarrow\infty$.
 So, from Eqs.~(\ref{E:PEC_spin_flip}) and~(\ref{E:PEC_inequality}) in
 Lemma~\ref{Lem:PEC}, the QER of the remaining quantum registers after PEC,
 $e^\textrm{final}$, is upper-bounded by
 \begin{eqnarray}
  & & e^\textrm{final} \nonumber \\
  & < & \frac{\epsilon_I}{2\ell} + (N-1) \times \nonumber \\
  & & ~~\exp \left\{ \frac{-\epsilon_I N (e_{00} -
   e_{01})^{2^{k+1}}}{8\ell (N-1) N^{2^k} e_{01}^{2^k} \left[ e_{00} + (N-1)
   e_{01} \right]^{2^k}} \right\} . \label{E:QER_p2}
 \end{eqnarray} 
 In other words, $e^\textrm{final} < \epsilon_I / \ell$ provided that
 \begin{equation}
  \left( e_{00} - e_{01} \right)^2 > N e_{01} \left[ e_{00} + (N-1) e_{01}
  \right] . \label{E:QER_Condition}
 \end{equation}
 This condition is satisfied if and only if 
 \begin{equation}
  e_{00} > \frac{N^2+1+(N^2-1)\sqrt{5}}{2N(N^2+N-1)} .
  \label{E:QER_Condition_e00}
 \end{equation}
 It is easy to verify that the constraint in Eq.~(\ref{E:QER_Condition_e00})
 is consistent with the assumption that $e_{00} > e_{01}$. Hence, provided
 that the initial QER satisfies
 \begin{equation}
  \sum_{(a,b)\neq (0,0)} e_{ab} < \frac{(N^2-1)(2N+1-\sqrt{5})}{2N(N^2+N-1)}
  = e^\textrm{QER} , \label{E:QER_Condition_QER}
 \end{equation}
 the fidelity of the $\ell$ quantum particle pairs shared between Alice and Bob
 immediately before they perform standard basis measurements to obtain their
 secret key is at least $1-e^\textrm{final} > 1-\epsilon_I/\ell$. By
 Footnote~28 in Ref.~\cite{lochauqkdsec}, the mutual information between Eve's
 final measurement result after eavesdropping and the final secret key is at
 most $\epsilon_I$. Thus, provided Alice and Bob abort the scheme if the
 estimated QER in step~\ref{Ent:QCTest} exceeds $(e^\textrm{QER}-\delta)$, the
 secret key generated is provably secure. That is to say, the scheme is
 unconditionally secure with security parameters $(\epsilon_p,\epsilon_I)$.
\end{proof}

\par\medskip
 A few remarks are in order. First, as Scheme~A reduces any kind of
 eavesdropping attacks in the channel to a classical cheating strategy which in
 turn is reduced to depolarization of the quantum signal, the ratio of the QER
 to the SBMER is given by $(N+1) : N$. From Theorem~\ref{Thrm:Uncond_Sec_A},
 the maximum tolerable SBMER for Scheme~A equals
\begin{equation}
 e^\textrm{SBMER} = \frac{(N^2-1)(2N+1-\sqrt{5})}{2(N+1)(N^2+N-1)} .
 \label{E:SBMER_QER}
\end{equation}
 In addition, if $p=2$, Lemma~\ref{Lem:Depolarize} implies that no matter what
 bijective map Alice and Bob use to convert their standard basis
 $2^n$-dimensional quantum particle measurement results into an $n$-bit string,
 the probability that exactly $i$ out of $n$ consecutive measured bits are in
 error equals $2^n e_{01} \binom{n}{i}$ for all $0\leq i\leq n$. Consequently,
 the BER equals $2^n e_{01} \sum_{i=0}^n \binom{n}{i} i /n = 2^{2n-1} e_{01}$;
 and the maximum tolerable BER for Scheme~A is given by
\begin{equation}
 e^\textrm{BER} = \frac{N(2N+1-\sqrt{5})}{4(N^2+N-1)} . \label{E:BER_QER}
\end{equation}
 We tabulate the tolerable SBMER and BER in Table~\ref{T:SBMER_BER}. However,
 we must emphasize once again that according to the discussions in
 Subsection~\ref{Subsec:Error_Rates}, we \emph{cannot} deduce the relative
 error tolerance capability from Table~\ref{T:SBMER_BER}.
 
\begin{table}
\begin{center}
\begin{tabular}{|r|c|c|} 
 \hline
 N & Tolerable SBMER & Tolerable BER \\ \hline
 2 & 27.64\% & 27.64\% \\
 3 & 43.31\% & N.A. \\
 4 & 53.40\% & 35.60\% \\
 5 & 60.44\% & N.A. \\
 7 & 69.62\% & N.A. \\
 8 & 72.78\% & 41.59\% \\
 9 & 75.34\% & N.A. \\
 11 & 79.25\% & N.A. \\
 13 & 82.09\% & N.A. \\
 16 & 85.14\% & 45.41\% \\
 \hline
\end{tabular}
\end{center}
\caption{The tolerable SBMER and BER for Scheme~A and hence also Schemes~B
 and~C for $N\leq 16$. As pointed out in the text, the values of SBMER and BER
 should not be compared directly.}
\label{T:SBMER_BER}
\end{table}
 
 Second, we study the tolerable error rate of Scheme~A as a function of $N$.
 Table~\ref{T:SBMER_BER} shows that the maximum tolerable BER $e^\textrm{BER}$
 for $N=2$ is the same as the one obtained earlier by Chau in
 Ref.~\cite{sixstateexact}. Additionally, $e^\textrm{SBMER}$ increases as $N$
 increases. In fact, the tolerable SBMER and BER tend to 100\% and 50\%
 respectively as $N\rightarrow\infty$. More precisely, as $n\rightarrow\infty$,
 the tolerable BER for Scheme~A using $2^n$-level quantum particles scales as
 $\approx 1/2 - (1+\sqrt{5})/2^{n+2}$. If $N$ is a prime power,
 $e^\textrm{SBMER}$ for Scheme~A using $N$-level quantum particles scales as
 $\approx 1-(3+\sqrt{5})/2N$ as $N\rightarrow\infty$. On the other hand, the
 lemma below sets the upper limit for the tolerable SBMER for Scheme~A.

\par\medskip
\begin{Lem}
 The tolerable SBMER for Scheme~A is upper-bounded by $(N-1)/(N+1)$. In fact,
 this bound is set by the following interpret-and-resend strategy: for each
 $N$-dimensional particle in the insecure quantum channel, Eve randomly and
 independently picks $M\in SL(2,N)$ and measures the particle in the basis $\{
 T(M)|i\rangle : i\in GF(N) \}$. Then, she records the measurement result and
 resends the measured particle to Bob. \label{Lem:UpperBound_QER_SBMER}
\end{Lem}
\begin{proof}
 The proof follows the idea reported in Ref.~\cite{qkd2waylocc}. Clearly, using
 this intercept-and-resend strategy, no quantum correlation between Alice and
 Bob can survive and hence no provably secure key can be distributed. Thus,
 this eavesdropping strategy sets the upper bound for the tolerable SMBER and
 BER for Scheme~A. If the quantum particle is prepared by Alice and measured by
 Eve in the same basis, that particle will suffer $Z_a$ error with equal
 probability for all $a\in GF(N)$. As Scheme~A depolarizes Pauli errors, we
 know that $e_{00}$ induced by this eavesdropping strategy equals $1/N$.
 Therefore, the SBMER for this strategy is $[(1-1/N)/(N^2-1)] \times N(N-1) =
 (N-1)/(N+1)$.
\end{proof}

\par\medskip\noindent
 Thus, the difference between the tolerable SBMER and its theoretical upper
 bound tends to zero in the limit of large $N$. So in this limit, the error
 tolerance capability of Scheme~A approaches its maximally allowable value.

 Third, readers may wonder why Scheme~A is highly error-tolerant especially
 when $N$ is large. Every quantum cheating strategy can be reduced to a
 classical one. Furthermore, Lemma~\ref{Lem:Depolarize} tells us that Scheme~A
 depolarizes the errors caused by any classical cheating strategy in the
 transmitted quantum signals. This greatly restricts the types of quantum
 errors we need to consider. The LOCC2 EP becomes a powerful tool to reduce
 spin errors at the expense of increasing phase errors. Furthermore,
 $e_{Z_0}^{k \,\textrm{EP}} > e_{Z_b}^{k \,\textrm{EP}}$ for all $b\neq 0$
 provided that $e_{00} > e_{01}$. In other words, the dominant kind of phase
 error is having no phase error at all. Thus, the majority vote PEC procedure
 is effective in bringing down the phase error. This is the underlying reason
 why Scheme~A is so powerful that, in the limit $N\rightarrow\infty$,
 $e^\textrm{SBMER}\rightarrow 1^-$.

 Fourth, the unconditional security proof in Theorem~\ref{Thrm:Uncond_Sec_A}
 does not depend on the phase $f_M(a,b)$ used in Eq.~(\ref{E:XZT}). Recall from
 the discussions in Subsection~\ref{Subsec:TConstruct} that $T: SL(2,2^n)
 \longrightarrow U(2^n)$ is not a group representation. So, in practice, Alice
 and Bob may replace $T(M_1 M_2 \cdots M_k)$ used in Scheme~A by $T(M_k)
 T(M_{k-1}) \cdots T(M_1)$, in which the $M_i$'s are chosen from the two
 generators of $SL(2,2^n)$.

 Fifth, the privacy amplification performed in Scheme~A is based entirely on
 entanglement purification and phase error correction. In fact, the key
 ingredient in reducing the QER used in the proof of
 Theorem~\ref{Thrm:Uncond_Sec_A} is the validity of the condition stated in
 Eq.~(\ref{E:QER_Condition}). Nonetheless, there is no need to bring down the
 QER to the small security parameter $\epsilon_I$. One may devise an equally
 secure scheme by following the adaptive procedure introduced by Chau in
 Ref.~\cite{sixstateexact} instead. That is to say, Alice and Bob may switch to
 a concatenated Calderbank-Shor-Steane quantum code when the PEC brings down
 the QER to about 5\%. The strategy of adding an extra step of quantum error
 correction towards the end of the privacy amplification procedure may increase
 the key generation rate. To understand why, let us consider the proof of
 Theorem~\ref{Thrm:Uncond_Sec_A} together with Eq.~(\ref{E:r_bound}). They tell
 us that in order to bring the QER down to less than $\epsilon$ after $k$
 rounds of LOCC2 EP, Alice and Bob have to choose $r$ and hence the number of
 quantum registers needed in PEC to be $\sim \epsilon c^{2^k}$ for some
 constant $c>1$. In contrast, by randomizing the quantum registers, the QER
 after each application of Steane's seven quantum register code is reduced
 quadratically whenever the QER is less than about 5\%. Consequently, Alice and
 Bob may increase the key generation rate by performing less rounds of
 LOCC2 EP, choosing $\epsilon \approx 0.01$, and finally adding a few rounds of
 the Calderbank-Shor-Steane code quantum error correction procedure.

\section{Reduction To The Prepare-And-Measure Scheme \label{Sec:ShorPre}}
 Finally, we apply the standard Shor and Preskill proof \cite{shorpre} to
 reduce the EB Scheme~A to two provably secure PM schemes in this section. Let
 us first write down the detail procedures of Schemes~B and~C before showing
 their security.
 
\par\medskip
\begin{enumerate}[\underline{PM QKD Scheme~B}]
 \item Alice randomly and independently prepares $L\gg 1$ quantum particles in
  the standard basis. She randomly and independently applies a unitary
  transformation $T(M) \in T[G]$ to each quantum particle, where $G$ equals
  $SL(2,N)$ or an order $(N^2-1)$ subgroup of $SL(2,N)$ (if it exists). Alice
  records the states and transformations she applied and then sends the states
  to Bob. Bob acknowledges the receipt of these particles and then applies a
  randomly and independently picked $T(M')^{-1}$ to each received particle.
  Now, Alice and Bob publicly reveal the unitary transformations they applied
  to each particle. A particle is kept and is said to be in the set $S_M$ if
  Alice and Bob have applied $T(M)$ and $T(M)^{-1}$ to it respectively. Bob
  measures the particles in $S_M$ in the standard basis and records the
  measurement results. \label{PM:Prepare}
 \item Alice and Bob estimate the channel quantum error rate by sacrificing a
  few particles. Specifically, they randomly pick $\mbox{O} ([N+1]^2 \log [
  \,|G| / \epsilon ] / \delta^2 N^2)$ pairs from each of the $|G|$ sets $S_M$
  and publicly reveal the preparation and measured states for each of them. In
  this way, they obtain the estimated channel error rate to within $\delta$
  with probability at least $(1-\epsilon)$. If the channel error rate is too
  high, they abort the scheme and start all over again. \label{PM:QCTest}
 \item Alice and Bob perform the following privacy amplification procedure.
   \begin{enumerate}
    \item They apply the privacy amplification procedure with two-way classical
     communication similar to the ones reported in
     Refs.~\cite{qkd2waylocc,sixstateexact}. Specifically, Alice and Bob
     randomly group their corresponding remaining quantum particles in pairs.
     Suppose the $j$th particle of the $i$th pair was initially prepared in the
     state $|s_{i_j}\rangle$. Then, Alice publicly announces the value $s_{i_1}
     -s_{i_2} \in GF(N)$ for each pair $i$. Similarly, Bob publicly announces
     the value $s'_{i_1}-s'_{i_2}$ where $|s'_{i_j}\rangle$ is the measurement
     result of the $j$th particle in the $i$th pair. They keep one of their
     corresponding registers of the pair only when their announced values of
     the corresponding pairs agree. They repeat the above procedure until there
     is an integer $r > 0$ such that a single application of
     step~\ref{Ent:PriAmp}b will bring the signal quantum error rate of the
     resultant particles down to $\epsilon_I /\ell^2$ for a fixed security
     parameter $\epsilon_I > 0$, where $r\ell$ is the number of remaining
     quantum particles they have. They abort the scheme either when $r$ is
     greater than the number of remaining quantum particles they possess or
     when they have used up all their quantum particles in this procedure.
    \item They apply the majority vote phase error correction procedure
     introduced by Gottesman and Lo \cite{qkd2waylocc}. Specifically, Alice and
     Bob randomly divide their corresponding resultant particles into sets
     each containing $r$ particles. They replace each set by the sum of the
     values prepared (by Alice) or measured (by Bob) of the $r$ particles in
     the set. These replaced values are bits of their final secure key string.
   \end{enumerate}
   \label{PM:PriAmp}
\end{enumerate}

\par\medskip\noindent
\begin{enumerate}[\underline{EQB PM QKD
 Scheme~C[$2^n,n_\textrm{\scriptsize ns}$]}]
 \item Alice and Bob agree on a bijection mapping $GF(2^n)$ to an $n$-bit
  string. Alice prepares $L\gg 1$ sets; and each set contains $n$ qubits that
  are randomly and independently prepared in the standard basis $\{ |i\rangle :
  i\in GF(2^n) \}$ identified through their mutually agreed bijection. She
  records the state of each set in the form of an $n$-bit string. Then, she
  randomly and independently applies $T(M) \in T[G]$ to each set of qubits,
  where $G$ equals $C_3 < SL(2,2)$ and $SL(2,2^n)$ for $n=1$ and $n>1$
  respectively. She permutes the $n$ qubits in each set with
  $n_\textrm{\scriptsize ns}$ randomly prepared non-signaling qubits and sends
  them to Bob. (In the upcoming analysis, one finds that for a fixed $n$, the
  tolerable BER of this scheme increases with $n_\textrm{\scriptsize ns}$.
  However, the number of non-signaling qubits used is limited by the absence of
  quantum storage capability.) After Bob has received these qubits, Alice tells
  him which of the $n$ qubits belong to a set that will be used to generate the
  key. Bob measures and discards the $n_\textrm{\scriptsize ns}$ non-signaling
  qubits and applies a randomly and independently picked $T(M')^{-1}$ to each
  of the $n$ qubits in the set that will be used to generate the key. Now,
  Alice and Bob publicly reveal their unitary transformations applied to each
  set. A set is kept and is said to be in $S_M$ if Alice and Bob have applied
  $T(M)$ and $T(M)^{-1}$ to it respectively. Bob records the standard basis
  measurement results identified through their mutually agreed on bijection in
  the form of an $n$-bit string for each set in $S_M$. At this point, Alice and
  Bob should each have $|G|$ families of $n$-bit strings; each family contains
  the prepare state/measurement result of qubits in $S_M$. Moreover, in the
  absence of noise and Eve, the corresponding bit strings in Alice's and Bob's
  hands should agree. \label{Qubit:Prepare}
 \item Alice and Bob regard their $|G|$ families of $n$-bit strings as states
  in the standard basis $\{ |i\rangle : i\in GF(N) \}$ and follow
  steps~\ref{PM:QCTest} and~\ref{PM:PriAmp} in Scheme~B to obtain their secret
  key. \label{Qubit:PriAmp}
\end{enumerate}

\par\medskip
 Note that in Scheme~C[$2^n,n_\textrm{\scriptsize ns}$] (or Scheme~C for short
 if the values of $n$ and $n_\textrm{\scriptsize ns}$ are clearly known to the
 readers), apart from the possibly entangled qubits that are used to generate
 the secret key, Alice and Bob have to create and send random non-signaling
 qubits through the insecure channel. The proofs of
 Theorems~\ref{Thrm:Uncond_Sec_B_C} and~\ref{Thrm:Advantage} below tell us that
 while the use of non-signaling qubits does not change the tolerable BER, it is
 essential for Scheme~C to tolerate more drastic eavesdropping attacks.

\par\medskip
\begin{Thrm}[Based on Shor and Preskill \cite{shorpre}]
 The tolerable BER of Scheme~A in Subsection~\ref{Subsec:EntScheme} as well as
 Schemes~B and~C above are equal. Thus, the conclusion of
 Theorem~\ref{Thrm:Uncond_Sec_A} is also applicable to Schemes~B and~C.
 \label{Thrm:Uncond_Sec_B_C} 
\end{Thrm}
\begin{proof} 
 Recall from Ref.~\cite{shorpre} that Alice may measure all her share of
 quantum registers right at step~\ref{Ent:Prepare} in Scheme~A without
 affecting the security of the scheme. Besides, LOCC2 EP and PEC procedures in
 Scheme~A simply permute the measurement basis. Also, the final secret key
 generation does not make use of the phase information of the transmitted
 quantum registers. Hence, the Shor-Preskill argument in Ref.~\cite{shorpre}
 can be applied to Scheme~A, giving us equally secure PM Schemes~B and~C. (Note
 that the introduction of random non-signaling qubits does not affect the
 tolerable BER of Scheme~C as these qubits are discarded after being measured
 and are not used to generate the secret key.)
\end{proof}

\par\medskip
 As discussed in Subsection~\ref{Subsec:Error_Rates}, we \emph{cannot} compare
 the error tolerant capability of Scheme~B that uses unentangled quantum
 particles of different dimensions as information carriers. Nonetheless, we can
 compare the error tolerant capability of the EQB PM QKD Scheme~C against the
 same eavesdropping attack.
 
\par\medskip
\begin{Thrm}
 For any fixed $n$, the error tolerant capability of
 Scheme~C[$2^n,n_\textrm{\scriptsize ns}$] increases with
 $n_\textrm{\scriptsize ns}$ in the limit of a large $\sum_{M\in SL(2,2^n)}
 |S_M|$. Besides, in the limits of a large $\sum_{M\in SL(2,2^n)} |S_M|$ and a
 large $n_\textrm{\scriptsize ns}$, the error tolerant capability of
 Scheme~C[$2^n,n_\textrm{\scriptsize ns}$] increases with $n$. That is to say,
 for any fixed $n$ and in the limit of a large $n_\textrm{\scriptsize ns}$,
 whenever Scheme~C[$2^n,n_\textrm{\scriptsize ns}$] generates a provably secure
 key under an eavesdropping attack, so does
 Scheme~C[$2^{n'},n_\textrm{\scriptsize ns}]$ under the same attack for any
 $n' > n$. Furthermore, there is a family of eavesdropping attacks that can be
 tolerated by Scheme~C[$2^{n'},n_\textrm{\scriptsize ns}$]. However, no
 provably secure key is produced in
 Scheme~C[$2^n,n_\textrm{\scriptsize ns}$]. \label{Thrm:Advantage}
\end{Thrm}
\begin{proof}
 Recall that Alice sends Bob packets of qubits each containing $n$ signaling
 as well as $n_\textrm{\scriptsize ns}$ non-signaling qubits and that any
 eavesdropping strategy in Scheme~C is equivalent to a classical probabilistic
 cheating strategy. Suppose that the channel quantum error rate is $q$. In
 other words, the probability that a randomly chosen qubit passing through the
 insecure channel is in error equals $q$. Let $q_k$ denote the portion of
 packets that contains exactly $k$ erroneous qubits. Then, $q_k$'s satisfy the
 following three constraints:
 \begin{equation}
  \sum_{k=0}^{n+n_\textrm{\tiny ns}} q_k = 1 , \label{E:q_k_1}
 \end{equation}
 \begin{equation}
  \sum_{k=0}^{n+n_\textrm{\tiny ns}} k \,q_k = (n+n_\textrm{\scriptsize ns})
  \,q , \label{E:q_k_2}
 \end{equation}
 and
 \begin{equation}
  0 \leq q_k \leq 1 \label{E:q_k_3}
 \end{equation}
 for $k = 0,1,\ldots ,n+n_\textrm{\scriptsize ns}$. Clearly, the set of $(q_0,
 q_1,\ldots, q_{n+n_\textrm{\tiny ns}})$ satisfying the above three constraints
 is convex.

 Since Eve does not know which qubits are signaling before Bob has received
 them, the QER for the signaling qubits is given by
 \begin{eqnarray}
  q_\textrm{\scriptsize QER} & = & \sum_{k=1}^{n+n_\textrm{\tiny ns}} \left( 1
   - \prod_{i=0}^{k-1} \frac{n_\textrm{\tiny ns}-i}{n+n_\textrm{\tiny ns}-i}
   \right) \,q_k \nonumber \\
  & = & 1 - \sum_{k=0}^{n+n_\textrm{\tiny ns}} \left( q_k \prod_{i=0}^{n-1}
   \frac{n+n_\textrm{\tiny ns}-k-i}{n+n_\textrm{\tiny ns}-i} \right) .
  \label{E:resultantQER}
 \end{eqnarray}
 We claim that for any $q_k$'s satisfying the three constraints
 (\ref{E:q_k_1})--(\ref{E:q_k_3}), $q_\textrm{\scriptsize QER}$ is
 upper-bounded by
 \begin{equation}
  q_\textrm{\scriptsize QER} \leq 1 - \sum_{k=\lfloor (n+n_\textrm{\tiny ns}) q
  \rfloor}^{\lfloor (n+n_\textrm{\tiny ns}) q \rfloor + 1} \left( \tilde{q}_k
  \prod_{i=0}^{n-1} \frac{n+n_\textrm{\tiny ns}-k-i}{n+n_\textrm{\tiny ns}-i}
  \right) , \label{E:resultantQER_bound}
 \end{equation}  
 where $\tilde{q}_k$'s are the (unique) solutions of the system of equations
 \begin{equation}
  \sum_{k=\lfloor (n+n_\textrm{\tiny ns}) q \rfloor}^{\lfloor (n+
  n_\textrm{\tiny ns}) q \rfloor + 1} \tilde{q}_k = 1 \label{E:tilde_q_1}
 \end{equation}
 and
 \begin{equation}
  \sum_{k=\lfloor (n+n_\textrm{\tiny ns}) q \rfloor}^{\lfloor (n+
  n_\textrm{\tiny ns}) q \rfloor + 1} k \,\tilde{q}_k = (n+
  n_\textrm{\scriptsize ns}) \,q . \label{E:tilde_q_2}
 \end{equation}
 In other words, we claim that among all strategies that cause a channel
 quantum error rate $q$, the one that causes either $\lfloor (n+
 n_\textrm{\scriptsize ns}) q \rfloor$ or $\lfloor (n+
 n_\textrm{\scriptsize ns}) q \rfloor + 1$ erroneous qubits in each packet
 produces the highest QER in the signaling qubits.

 To show the validity of our claim, we rewrite Eq.~(\ref{E:resultantQER}) as
 \begin{eqnarray}
  q_\textrm{\scriptsize QER} & = & 1 - \sum_{k=\lfloor (n+n_\textrm{\tiny ns})
   q \rfloor}^{\lfloor (n+n_\textrm{\tiny ns}) q \rfloor + 1} \left(
   \tilde{q}_k \prod_{i=0}^{n-1} \frac{n+n_\textrm{\tiny ns}-k-i}{n+
   n_\textrm{\tiny ns}-i} \right) \nonumber \\
  & & ~- \sum_{k=0}^{n+n_\textrm{\tiny ns}} \left( \Delta q_k \prod_{i=0}^{n-1}
   \frac{n+n_\textrm{\tiny ns}-k-i}{n+n_\textrm{\tiny ns}-i} \right) ,
  \label{E:resultantQER_rewritten}
 \end{eqnarray}
 where $\Delta q_k = q_k - \tilde{q}_k$ if $k = \lfloor (n+
 n_\textrm{\scriptsize ns}) q \rfloor$ or $\lfloor (n+
 n_\textrm{\scriptsize ns}) q \rfloor + 1$, and $\Delta q_k = q_k$ otherwise.
 Since the set of $(q_0,\ldots ,q_{n+n_\textrm{\tiny ns}})$ satisfying
 Eqs.~(\ref{E:q_k_1})--(\ref{E:q_k_3}) is convex, the claim is valid if we can
 show that the last term in Eq.~(\ref{E:resultantQER_rewritten}) is
 non-positive for all $\Delta q_k$'s satisfying $\sum_k \Delta q_k = \sum_k k
 \,\Delta q_k = 0$ and $\Delta q_j \geq 0$ whenever $j = \lfloor (n+
 n_\textrm{\scriptsize ns}) q \rfloor$ or $\lfloor (n+
 n_\textrm{\scriptsize ns}) q \rfloor + 1$.

 There are three cases to consider. The first case is that $\Delta q_k \geq 0$
 for all $k$. Clearly, this is possible only if $\Delta q_k = 0$ for all $k$.
 So in this case, the last term in Eq.~(\ref{E:resultantQER_rewritten}) equals
 $0$.

 The second case is that exactly one $\Delta q_k < 0$. Without lost of
 generality, we may assume that the one is $\Delta q_{\lfloor (n+
 n_\textrm{\tiny ns}) q \rfloor}$. Observe that one can tune $\Lambda_i$'s to
 make the auxiliary real-valued function $\xi$ in the equation below two times
 differentiable and $\xi'' \geq 0$ in $(0,n+n_\textrm{\scriptsize ns})$:
 \begin{equation}
  \xi (k) = \left\{ \begin{array}{ll} \displaystyle \prod_{i=0}^{n-1} \Gamma
  (n+n_\textrm{\scriptsize ns}-k-i+1) & \mbox{if~} 0\leq k\leq
  n_\textrm{\scriptsize ns}, \\
  \\
  \displaystyle \sum_{i=0}^3 \Lambda_i k^i & \mbox{if~}
   n_\textrm{\scriptsize ns} < k < n_\textrm{\scriptsize ns} + 1, \\
  \\
  0 & \mbox{if~} k \geq n_\textrm{\scriptsize ns} + 1 . \end{array} \right.
  \label{E:generalized_dist}
 \end{equation}
 Consequently, such a $\xi (k)$ is a convex function in the interval $[0,n+
 n_\textrm{\scriptsize ns}]$. Since $\sum_{k\neq \lfloor (n+
 n_\textrm{\tiny ns}) q \rfloor} \Delta q_k = - \Delta q_{\lfloor (n+
 n_\textrm{\tiny ns}) q \rfloor} > 0$, the convexity of $\xi$ implies that the
 last term in Eq.~(\ref{E:resultantQER_rewritten}) is non-positive.

 The last case is that exactly two $\Delta q_k < 0$, namely, for $k = \lfloor
 (n+n_\textrm{\scriptsize ns}) q \rfloor$ and $\lfloor (n+
 n_\textrm{\scriptsize ns}) q \rfloor + 1$. In this situation, $\sum_k \Delta
 q_k = \sum_k k \,\Delta q_k = 0$  demands that there exist $\Delta q_{k_1},
 \Delta q_{k_2} > 0$ for some $k_1 < \lfloor (n+n_\textrm{\scriptsize ns}) q
 \rfloor$ and $k_2 > \lfloor (n+n_\textrm{\scriptsize ns}) q \rfloor + 1$.
 Consequently, we may define $\Delta q'_j = \Delta q''_j = 0$ for $j = \lfloor
 (n+n_\textrm{\scriptsize ns}) q \rfloor$ and $\lfloor (n+
 n_\textrm{\scriptsize ns}) q \rfloor + 1$ and decompose $\Delta q_k$ as
 $\Delta q'_k + \Delta q''_k$ for all $k \neq \lfloor (n+
 n_\textrm{\scriptsize ns}) q \rfloor$, $\lfloor (n+n_\textrm{\scriptsize ns})
 q \rfloor + 1$ in such a way that $\Delta q'_k, \Delta q''_k \geq 0$ for all
 $k$ and $\sum_k \Delta q'_k = - \Delta q_{\lfloor (n+n_\textrm{\tiny ns}) q
 \rfloor}$ and $\sum_k \Delta q''_k = - \Delta q_{\lfloor (n+
 n_\textrm{\tiny ns}) q \rfloor + 1}$. By means of this decomposition and the
 convexity of the function $\xi$, we conclude that the last term in
 Eq.~(\ref{E:resultantQER_rewritten}) is non-positive. Hence, the claim in
 Eq.~(\ref{E:resultantQER_bound}) is valid.

 From Eqs.~(\ref{E:resultantQER_bound})--(\ref{E:tilde_q_2}), it is easy to
 check that for a fixed $n$, tolerable BER of
 Scheme~C[$2^n,n_\textrm{\scriptsize ns}$] increases with
 $n_\textrm{\scriptsize ns}$. Combining with Eq.~(\ref{E:SBMER_QER}) and
 Table~\ref{T:SBMER_BER}, we conclude that for $n=2$, $q \approx 1.5\times
 0.2764$ and $q_\textrm{\scriptsize QER} \lesssim 1.25\times 0.5340$,
 $n_\textrm{\scriptsize ns} \geq 23$. Thus,
 Scheme~C[$4,n_\textrm{\scriptsize ns}$] generates a provably secure key when
 the channel bit error rate is slightly higher than 27.64\% provided that
 $n_\textrm{\scriptsize ns} \geq 23$. Thus, this scheme is more error-resistant
 than any UQB QKD scheme known to date.

 Note that as $n_\textrm{\scriptsize ns} \rightarrow \infty$, the right hand
 side of Eq.~(\ref{E:resultantQER_bound}) becomes $1-(1-q)^n$. (A simple way to
 argue why this is the case is to observe that in the limit of a large number
 of random non-signaling qubits used, Eve can do no better than guessing which
 of the $n$ qubits in a packet are used to generate the secret key when these
 qubits are traveling in the insecure channel.) As the Pauli signal quantum
 error is depolarized, Lemma~\ref{Lem:Depolarize} demands that the error rates
 caused by this classical probabilistic strategy are given by
 \begin{equation}
  e_{ab} = \left\{ \begin{array}{cl} (1-q)^n & \mbox{if~} a=b=0, \\ ~ \\
   \displaystyle \frac{1-(1-q)^n}{2^{2n}-1} & \mbox{otherwise.} \end{array}
  \right. \label{E:error_rate_random_q}
 \end{equation}
 From Eq.~(\ref{E:BER_QER}), the final key is provably secure provided that the
 probability $q$ satisfies
 \begin{equation}
  q < q_\textrm{\scriptsize crit} (n) \equiv 1-\frac{1}{2} \left[ \frac{(1+
  \sqrt{5}) 2^{2n} - (\sqrt{5}-1)}{2(2^{2n}+2^n-1)} \right]^{1/n} .
  \label{E:q_bound}
 \end{equation}
 Since $q_\textrm{\scriptsize crit} (n)$ is a strictly increasing function of
 $n$, we conclude that the error tolerant capability of
 Scheme~C[$2^n,n_\textrm{\scriptsize ns}$] strictly increases with increasing
 $n$ in the limit of large $n_\textrm{\scriptsize ns}$. Hence, this theorem is
 proved.
\end{proof}

\par\medskip
 Since the most error-resistant UQB PM scheme known to date is the one offered
 by Chau in Ref.~\cite{sixstateexact} (which is also equivalent to
 Scheme~C[$2,0$]), the above theorem clearly shows the advantage of using
 entangled qubits as information carriers provided that Alice and Bob can
 transmit a large number of qubits without requiring quantum storage.
 Specifically, no UQB PM scheme to date can generate a provably secure key if
 Eve randomly causes an error to a qubit in the insecure quantum channel with
 probability $q$ satisfying $0.4146 \approx q_\textrm{\scriptsize crit} (1)
 \leq q < q_\textrm{\scriptsize crit} (2) \approx 0.4234$. In contrast,
 Scheme~C[$2^n,n_\textrm{\scriptsize ns}$] tolerates such an attack for any $n
 \geq 2$ and for a sufficiently large $n_\textrm{\scriptsize ns}$ depending on
 $n$.

 We emphasize that the use of random non-signaling qubits is vital in the proof
 of Theorem~\ref{Thrm:Advantage}. Otherwise, Eve may cause 100\% signal quantum
 error in Scheme~C[$2^n,n_\textrm{\scriptsize ns}$] by creating an $X$ error to
 every one out of $n$ consecutive qubits that passes through the insecure
 quantum channel. However, we also have to stress that the presence of
 non-signaling qubits lowers the key generation rate of Scheme~C. In the
 absence of quantum storage, the number of non-signaling qubits per packet
 $n_\textrm{\scriptsize ns}$ is limited by the decoherence time of qubits and
 the qubit transmission rate in the channel. The proof of
 Theorem~\ref{Thrm:Advantage} tells us that for $n=2$, Alice and Bob need to
 use $n_\textrm{\scriptsize ns} = 23$ in order to generate a provably secure
 key at a channel BER slightly higher than that which can be tolerated by all
 UQB QKD schemes known to date. Clearly, Scheme~C[$4,23$] generates a key at a
 rate 8\% that of Scheme~C[$2,0$]. Moreover, manipulating a packet of 25 qubits
 in the absence of quantum storage in Scheme~C[$4,23$] is challenging.
 
 Now, we discuss the number of different kinds of states Alice and Bob have to
 prepare and measure in Schemes~B and~C.

\par\medskip
\begin{Thrm}
 Suppose Alice and Bob follow Schemes~B or~C with $G = SL(2,N)$, so that they
 prepare and measure in $N(N+1)$ bases (and hence $N^2(N+1)$ different states).
 If they choose $G$ to be an order $(N^2-1)$ subgroup of $SL(2,N)$ instead,
 they need to prepare and measure in $(N+1)$ different bases (and hence $N(N+
 1)$ states). \label{Thrm:num_of_bases}
\end{Thrm}
\begin{proof}
 Case~(1): $G = SL(2,N)$. Let $G'$ be the subgroup $\{ \textrm{diag} (\alpha,
 \alpha^{-1}) : \alpha\in GF(N)^* \}$ of $SL(2,N)$. Let $g,g' \in G'$ and $h\in
 SL(2,N)$. From Eqs.~(\ref{E:XZT})--(\ref{E:phase_factor}), $\langle i| T(g
 h)^{-1} T(g' h) |i'\rangle$ $=\omega_p^{-\textrm{Tr} (i' k)} \langle i| T(g
 h)^{-1} T(g' h) Z_k |i'\rangle$ $=\omega_p^{-\textrm{Tr} (i' k)} \langle i|
 Z_{k\beta^{-1}} T(g h)^{-1} T(g' h) |i'\rangle$ $=\omega_p^{\textrm{Tr} ([
 \beta^{-1} i - i']k)}$ $\langle i| T(g h)^{-1} T(g' h) |i'\rangle$ for all $k
 \in GF(N)$, where $g' g = \textrm{diag} (\beta,\beta^{-1})$. Therefore,
 $\langle i| T(g h)^{-1} T(g' h) |i'\rangle  = 0$ if $i\neq i'\beta$. In other
 words, the bases $\{ T(g h) |i\rangle : i\in GF(N) \}$ and $\{ T(g' h) |i
 \rangle : i\in GF(N) \}$ are the same. Consequently, if Alice and Bob choose
 $G = SL(2,N)$ in Schemes~B and~C, they need to prepare and measure in $N(N^2-
 1)/(N-1) = N(N+1)$ bases (and hence $N^2(N+1)$ different states).
 
 Case~(2): $N=2$ and $G$ is the order $3$ subgroup of $SL(2,2)$.
 Theorem~\ref{Thrm:Subgroup} in the Appendix tells us that $G$ is unique. It
 is clear that, in this case, Alice and Bob need to prepare and measure their
 quantum states in three different bases.
 
 Case~(3): $N>2$ and $G$ is the order $(N^2-1)$ subgroup of $SL(2,N)$.
 Theorem~\ref{Thrm:Subgroup} in the Appendix implies that $N=3,5,7,11$.
 Besides, $G$ contains an order $(N-1)$ subgroup $H'$ in the form $\{ P^{-1}
 \textrm{diag} (\alpha,\alpha^{-1}) P : \alpha\in GF(N)^* \}$ for some $P\in
 SL(2,N)$. Recall from Subsection~\ref{Subsec:TConstruct} that $T : SL(2,N)
 \longrightarrow U(N)$ in this case is a transposed representation. Hence,
 from Eq.~(\ref{E:phase_factor}), $\langle i| T(g h)^{-1} T(g' h) |i'\rangle$
 $=\langle i| T(g' g^{-1}) |i'\rangle$ $=\langle i| T(\textrm{diag} (\beta,
 \beta^{-1})) |i'\rangle$ $=\langle i| i' \beta^{-1} \rangle$ for some $\beta
 \in GF(N)^*$. Hence, Alice and Bob need to prepare and measure in $(N^2-1) /
 (N-1) = N+1$ different bases (and hence $N(N+1)$ states).
\end{proof}

\par\medskip
 Since the maximum number of mutually unbiased bases equals $(N+1)$ for any
 prime power $N$ \cite{mub1,mub2,mub3}, Scheme~B shows that certain PM QKD
 schemes not using mutually unbiased bases can be more error-tolerant.

\section{Discussions \label{Sec:Dis}}
 In summary, we have introduced two PM QKD schemes (Schemes~B and~C) based on
 depolarization of Pauli errors and proved their unconditional security. In
 particular, we showed that for a sufficiently large Hilbert space dimension of
 quantum particles $N$ used, Scheme~B generates a provably secure key close to
 100\% SBMER or 50\% BER. This result demonstrates the advantages of using
 unentangled higher dimensional quantum particles as signal carriers as well as
 depolarizing Pauli errors in QKD. It also shows that, for $N>2$, the use of
 certain non-mutually unbiased bases increases the error tolerance capability
 of QKD. In addition, Scheme~C shows that the ability to create and transfer,
 but not to store entangled qubits is advantageous in quantum cryptography.

 There is a tradeoff between the error tolerance rate and key generation
 efficiency, however. It is clear from the proof of
 Theorem~\ref{Thrm:Uncond_Sec_A} that $r$, and hence the number $L$ of quantum
 particles transferred from Alice and Bob, scales as $2^k$. Besides, the
 probability that the measurement results agree and hence the control quantum
 register pairs are kept in LOCC2 EP equals $\approx 1/N$ in the worst case.
 As a result, while Schemes~B and~C are highly error-tolerant, they generate a
 secret key with exponentially small efficiency in the worst case scenario.
 Fortunately, the adaptive nature of Schemes~B and~C makes sure that this
 scenario will not happen when the error rate of the channel is small. To
 conclude, in most practical situations, Alice and Bob should choose the
 smallest possible $N$ whose corresponding $e^\textrm{SBMER}$ is slightly
 larger than the channel standard basis measurement error rate. In this way,
 they can generate their provably secure key at the highest possible rate.

\appendix
\section{Appendix}
 This appendix discusses the possibility of depolarizing Pauli error using
 proper subgroups of $SL(2,N)$. The analysis makes use of the Dickson theorem
 \cite{Dickson} on the subgroup classification of projective special linear
 groups over finite fields. The version of the Dickson theorem listed below is
 due to Huppert in the Hauptsatz~8.27 in Ref.~\cite{Huppert}.

\par\medskip
\begin{Thrm}[Dickson]
 Let $N=p^n$. Subgroups of $PSL(2,N)$ are isomorphic to one of the following
 families of groups:
 \begin{enumerate}
  \item Elementary Abelian $p$-groups;
  \item Cyclic groups $C_z$ of order $z$, where $z$ is a divisor of $(N\pm 1)/
   (N-1,2)$ \label{dickson_z};
  \item Dihedral groups $D_z$ of order $2z$, where $z$ is as defined in
   \ref{dickson_z});
  \item Alternating group $A_4$ (this can occur only for $p>2$ or when $p=2$
   and $n \equiv 0 \bmod 2$);
  \item Symmetric group $S_4$ (this can occur only if $N^2 \equiv 1 \bmod 16$);
  \item Alternating group $A_5$ (this can occur only if $p=5$ or $N^2 \equiv 1
   \bmod 5$);
  \item A semidirect product of an elementary Abelian group of order $p^m$ with
   a cyclic group of order $t$, where $t$ is a divisor of $(p^m - 1,N-1)$;
  \item The group $PSL(2,p^m)$ for $m$ a divisor of $n$, or the group
   $PGL(2,p^m)$ for $2m$ a divisor of $n$.
 \end{enumerate}
 \label{Thrm:Dickson}
\end{Thrm}

\par\medskip
 In addition to the Dickson theorem, the following lemma is also needed.

\par\medskip
\begin{Lem}
 If $N$ is odd, $-I$ is the only element in $SL(2,N)$ whose order is $2$.
 \label{Lem:order2_element}
\end{Lem}
\begin{proof}
 Let $M = \left[ \begin{array}{cc} \alpha & \beta \\ \delta & \gamma
 \end{array} \right]$ be an order $2$ element in $SL(2,N)$. $M^2 = I$ implies
 $\beta (\alpha + \gamma) = \delta (\alpha + \gamma) = 0$ and $\alpha^2 + \beta
 \delta = 1$. If $\alpha+\gamma = 0$, $\det M = -\alpha^2 - \beta\delta = 1$ is
 consistent with $\alpha^2 + \beta\delta = 1$ only if $N$ is even. So, $\alpha
 +\gamma$ must be equal to $0$. Hence, $\beta=\gamma=0$ and $M = \pm I$. As $N$
 is odd, $-I$ is the only order $2$ element in $SL(2,N)$.
\end{proof}

\par\medskip
 We examine the possibility of using a smaller group to depolarize Pauli error
 in step~\ref{Ent:Prepare}. Specifically, we look for subgroups $H$ of
 $SL(2,N)$ to do the job. Clearly, the order of the subgroup $H$ must be a
 multiple of $(N^2-1)$.

\par\medskip
\begin{Thrm}
 Proper subgroups $H$ of $SL(2,N)$ satisfying $(N^2-1) \mid |H|$ exist only for
 $N=2,3,5,7,11$ and $|H| = N^2-1$. Specifically,
 \begin{enumerate}
  \item When $N=2$, $H \cong C_3$. Moreover, this subgroup is unique and is
   generated by one element. In fact, $H = \left< \left[ \begin{array}{cc} 0 &
   1 \\ 1 & 1 \end{array} \right] \right>$.
  \item When $N=3$, $H \cong Q_8$. Moreover, this subgroup is unique and is
   generated by two elements. In fact, $H = \left< \left[ \begin{array}{cc} 1 &
   1 \\ 1 & 2 \end{array} \right] , \left[ \begin{array}{cc} 1 & 2 \\ 2 & 2
   \end{array} \right] \right>$.
  \item When $N=5$, $H/\{\pm I\} \cong A_4$. Moreover, $H$ is generated by
   two elements. One possible choice of $H$ is $\left< \left[ \begin{array}{cc}
   2 & 0 \\ 0 & 3 \end{array} \right] , \left[ \begin{array}{cc} 1 & 2 \\ 1 & 3
   \end{array} \right] \right>$.
  \item When $N=7$, $H/\{\pm I\} \cong S_4$. Moreover, $H$ is generated by
   two elements. One possible choice of $H$ is $\left< \left[ \begin{array}{cc}
   2 & 0 \\ 0 & 4 \end{array} \right] , \left[ \begin{array}{cc} 1 & 2 \\ 1 & 3
   \end{array} \right] \right>$.
  \item When $N=11$, $H/\{\pm I\} \cong A_5$. Moreover, $H$ is generated by
   two elements. One possible choice of $H$ is $\left< \left[ \begin{array}{cc}
   2 & 0 \\ 0 & 6 \end{array} \right] , \left[ \begin{array}{cc} 1 & 1 \\ 1 & 2
   \end{array} \right] \right>$.
 \end{enumerate}
 Furthermore, $| \{ M\in H : M [a~b]^t = [c~d]^t \} | = 1$ for all $[a~b],
 [c~d] \neq [0~0]$. Thus, replacing $SL(2,N)$ by $H$ in Scheme~A also
 depolarizes Pauli errors. \label{Thrm:Subgroup}
\end{Thrm}
\begin{proof}
 From the Dickson theorem, it follows that $SL(2,N)$ does not contain a proper
 subgroup $H$ whose order divides $(N^2-1)$ if $N\neq 2,3,5,7,11$. Moreover, if
 $H$ exists for $N=2,3,5,7,11$, $|H| = N^2-1$. In what follows, we are going to
 show that such $H$ indeed exist for $N=2,3,5,7,11$.

 Case~(1): When $N=2$, the Dickson theorem implies that if $H$ exists, $H \cong
 C_3$. Since the only order $3$ elements of $SL(2,2)$ are $M_{21} \equiv \left[
 \begin{array}{cc} 0 & 1 \\ 1 & 1 \end{array} \right]$ and $M_{21}^2$, the
 order $3$ subgroup $H$ of $SL(2,2)$ exists and is unique. An explicit
 expression for $T(M_{21})$ is given in Table~\ref{T:T} for reference.

 Case~(2): When $N=3$, the Dickson theorem implies that if $H$ exists, $H / \{
 \pm I \} \cong D_2 \cong C_2 \times C_2$. $H$ cannot be Abelian as $H$ would
 then be isomorphic to $C_2 \times C_2 \times C_2$, contradicting
 Lemma~\ref{Lem:order2_element}. Since $H$ is a non-Abelian group of order $8$,
 $H$ is generated by two elements. By Lemma~\ref{Lem:order2_element} and the
 proof of Proposition~6.3 in Ref.~\cite{Hungerford}, we conclude that the two
 elements generating $H$ are both of order $4$. Hence, $H \cong Q_8$. Note that
 the only order $4$ elements of $SL(2,3)$ are $M_{31} \equiv \left[
 \begin{array}{cc} 1 & 1 \\ 1 & 2 \end{array} \right]$, $-M_{31}$, $M_{32}
 \equiv \left[ \begin{array}{cc} 1 & 2 \\ 2 & 2 \end{array} \right]$,
 $-M_{32}$, $M_{31} M_{32}$ and $M_{32} M_{31}$. Therefore, $\left< M_{31},
 M_{32} \right>$ is the only order $(N^2-1)$ subgroup of $SL(2,3)$. Explicit
 expressions for $T(M_{31})$ and $T(M_{32})$ are given in Table~\ref{T:T} for
 reference.

 Case~(3): When $N=5$, the Dickson theorem implies that if $H$ exists, $H / \{
 \pm I \} \cong A_4$ or $D_6$. Satz~8.13 in Ref.~\cite{Huppert} says that
 $PSL(2,5) \cong A_5$. Hence, the only possibility is that $H / \{ \pm I \}
 \cong A_4$.
 Since $A_4$ can be generated by two elements, one of order $2$ and the other
 of order $3$, $H / \{ \pm I \} = \left< M_{51}/\{ \pm I \}, M_{52}/\{ \pm I \}
 \right>$ for some $M_{51}, M_{52} \in SL(2,5)$ provided that $H$ exists.
 Moreover $M_{51}/\{ \pm I \}$ and $M_{52}/\{ \pm I \}$ are of order $2$ and
 $3$, respectively. We may assume that $M_{52}^3 = -I$, for otherwise replace
 $M_{52}$ by $-M_{52}$.  Consequently, the subgroup $H$, if it exists, is equal
 to $\left< -I,M_{51},M_{52} \right> = \left< M_{51},M_{52} \right>$. Thus, $H$
 can be generated by two elements in $SL(2,5)$. From
 Lemma~\ref{Lem:order2_element}, the order of $M_{51}$ is equal to $4$. By
 explicit search, $H$ exists but is not unique. One possible $H$ is $\left\{
 \left[ \begin{array}{cc} 2 & 0 \\ 0 & 3 \end{array} \right], \left[
 \begin{array}{cc} 1 & 2 \\ 1 & 3 \end{array} \right] \right\}$.

 Case~(4): When $N=7$, the Dickson theorem implies that if $H$ exists, $H / \{
 \pm I \} \cong S_4$. Since $S_4$ is generated by two elements, namely
 $(1234)$ and $(123)$, the subgroup $H/\{ \pm I \}$, if it exists, equals
 $\left< M_{71} / \{ \pm I \} , M_{72} / \{ \pm I \} \right>$. Moreover, using
 the same argument as in the proof of case~(3), we may choose $M_{71}^4 = \pm
 I$ and $M_{72}^3 = -I$. Hence, $H$, if it exists, is equal to $\left< -I,
 M_{71}, M_{72} \right> = \left< M_{71}, M_{72} \right>$. By an explicit
 search, $H$ exists but is not unique. One possible $H$ is $\left< \left[
 \begin{array}{cc} 2 & 0 \\ 0 & 4 \end{array} \right] , \left[
 \begin{array}{cc} 1 & 2 \\ 1 & 3 \end{array} \right] \right>$.

 Case~(5): When $N=11$, the Dickson theorem implies that if $H$ exists, $H / \{
 \pm I \} \cong A_5$. Since $A_5$ is generated by two elements, namely
 $(12345)$ and $(123)$, using the same argument as in the proof of cases~(3)
 and~(4), we conclude that $H$, if it exists, can be generated by two elements.
 An explicit search tells us that $H$ exists but not unique, and one possible
 $H$ is $\left< \left[ \begin{array}{cc} 2 & 0 \\ 0 & 6 \end{array} \right] ,
 \left[ \begin{array}{cc} 1 & 1 \\ 1 & 2 \end{array} \right] \right>$.

 To show that $|\{ M\in H: M [a~b]^t = [c~d]^t \}| = 1$ for all $[a~b], [c~d]
 \neq [0~0]$, we observe from our discussion of the structure of $H$ above,
 that $H$ contains an order $(N-1)$ proper subgroup $H'$. Since $H' < SL(2,N)$,
 $H' = \{ P^{-1} \textrm{diag}(\alpha,\alpha^{-1}) P : \alpha\in GF(N)^* \}$
 for some $P\in SL(2,N)$. As all order $(N^2-1)$ subgroups of $SL(2,N)$ are
 conjugate to each other, it suffices to show the validity for $P = I$. As $N
 \nmid |H| = N^2-1$, $H$ does not contain elements of the form $\left[
 \begin{array}{cc} \alpha & \beta \\ 0 & \alpha^{-1} \end{array} \right]$ or
 $\left[ \begin{array}{cc} 0 & \alpha \\ -\alpha^{-1} & \beta \end{array}
 \right]$ for some $\beta \neq 0$. Therefore, for any $M = \left[
 \begin{array}{cc} \alpha & \beta \\ \delta & \gamma \end{array} \right] \in
 SL(2,N)$,
 \begin{eqnarray}
  |\{ H' M H' \}| & = & |\{ M' M M'' : M',M'' \in H' \}| \nonumber \\
  &  = & \left\{ \begin{array}{cl} N-1 & \textrm{if~} \alpha = 0
  \textrm{~or~} \delta = 0, \\ (N-1)^2 & \textrm{if~} \alpha,\delta \neq 0.
  \end{array} \right. \label{E:num_conj}
 \end{eqnarray}
 Also, the first column of matrices in $H' M H'$ are all distinct. Since $|H| =
 N^2-1$, Eq.~(\ref{E:num_conj}) requires that the first columns of the matrices
 in $H$ are all distinct. Hence, $|\{ M\in H : M [a~b]^t = [c~d]^t \}| = 1$ for
 all $[a~b], [c~d] \neq [0~0]$. Combining with the fact that $H'$ is a group,
 Scheme~A depolarizes Pauli errors.
\end{proof}

\section*{Acknowledgments}
 This work is supported in part by the RGC grant HKU~7010/04P of the HKSAR
 government.
\bibliographystyle{ieeetr}
\bibliography{qc33.3}

\begin{thebibliography}{10}

\bibitem{bbjm}
C.~H. Bennett, G.~Brassard, R.~Jozsa, D.~Mayers, A.~Peres, B.~Schumacher, and
  W.~K. Wootters, ``Reduction of quantum entropy by reversible extraction of
  classical information,'' {\em J.\ Mod.\ Opt.}, vol.~41, pp.~2307--2314, 1994.

\bibitem{biasedbb84}
H.-K. Lo, H.~F. Chau, and M.~Ardehali, ``Efficient quantum key distribution
  scheme and proof of its unconditional security,'' 2001.
\newblock (quant-ph/0011056v2), to appear in J.\ Crypt.

\bibitem{lochauqkdsec}
H.-K. Lo and H.~F. Chau, ``Unconditional security of quantum key distribution
  over arbitrarily long distances,'' {\em Science}, vol.~283, pp.~2050--2056,
  1999.
\newblock As well as the supplementary material available at
  {\texttt{http://www.sciencemag.org/feature/data/984035.shl}}.

\bibitem{mayersjacm}
D.~Mayers, ``Unconditional security in quantum cryptography,'' {\em J.\ Assoc.\
  Comp.\ Mach.}, vol.~48, pp.~351--406, 2001.
\newblock See also his preliminary version in D. Mayers, \textit{Advances in
  Cryptology --- Proceedings of Crypto'96} (Springer Verlag, Berlin, 1996),
  pp.~343--357.

\bibitem{gottloreview}
D.~Gottesman and H.-K. Lo, ``From quantum cheating to quantum security,'' {\em
  Phys.\ Today}, vol.~53, no.~11, pp.~22--27, 2000.
\newblock And references cited therein.

\bibitem{gisinreview}
N.~Gisin, G.~Ribordy, W.~Tittel, and H.~Zbinden, ``Quantum cryptography,'' {\em
  Rev.\ Mod.\ Phys.}, vol.~74, pp.~145--195, 2002.
\newblock And references cited therein.

\bibitem{bb84}
C.~H. Bennett and G.~Brassard, ``Quantum cryptography: Public key distribution
  and coin tossing,'' in {\em Proceedings of the IEEE International Conference
  on Computers, Systems and Signal Processing}, (New York), pp.~175--179,
  Bangalore, India, IEEE, 1984.

\bibitem{sixstate}
D.~Bru{\ss}, ``Optimal eavesdropping in quantum cryptography with six states,''
  {\em Phys.\ Rev.\ Lett.}, vol.~81, pp.~3018--3021, 1998.

\bibitem{cont1}
T.~C. Ralph, ``Continuous variable quantum cryptography,'' {\em Phys.\ Rev.\
  A}, vol.~61, pp.~010303(R):1--4, 2000.

\bibitem{cont2}
M.~Hillery, ``Quantum cryptography with sequeezed states,'' {\em Phys.\ Rev.\
  A}, vol.~61, pp.~022309:1--8, 2000.

\bibitem{squeez}
D.~Gottesman and J.~Preskill, ``Secure quantum key distribution using squeezed
  states,'' {\em Phys.\ Rev.\ A}, vol.~63, pp.~022309:1--18, 2001.

\bibitem{threestate}
H.~Bechmann-Pasquinucci and A.~Peres, ``Quantum cryptography with 3-state
  systems,'' {\em Phys.\ Rev.\ Lett.}, vol.~85, pp.~3313--3316, 2000.

\bibitem{alph}
H.~Bechmann-Pasquinucci and W.~Tittel, ``Quantum cryptography using larger
  alphabets,'' {\em Phys.\ Rev.\ A}, vol.~61, pp.~062308:1--6, 2000.

\bibitem{highdim}
M.~Bourennane, A.~Karlsson, and G.~Bj{\"o}rk, ``Quantum key distribution using
  multilevel encoding,'' {\em Phys.\ Rev.\ A}, vol.~64, pp.~012306:1--5, 2001.

\bibitem{dlevel}
N.~J. Cerf, M.~Bourennane, A.~Karlsson, and N.~Gisin, ``Security of quantum key
  distribution using d-level systems,'' {\em Phys.\ Rev.\ Lett.}, vol.~88,
  pp.~127902:1--4, 2002.

\bibitem{highdimmore}
M.~Bourennane, A.~Karlsson, G.~Bj{\"o}rk, N.~Gisin, and N.~J. Cerf, ``Quantum
  key distribution using multilevel encoding: \uppercase{S}ecurity analysis,''
  {\em J.\ Phys.:\ A}, vol.~35, pp.~10065--10076, 2002.

\bibitem{earlier_version}
H.~F. Chau, ``Unconditionally secure key distribution in higher dimensions,''
  2004.
\newblock (quant-ph/0212055v2).

\bibitem{evethreestate}
D.~Bru{\ss} and C.~Macchiavello, ``Optimal eavesdropping in cryptography with
  three-dimensional quantum states,'' {\em Phys.\ Rev.\ Lett.}, vol.~88,
  pp.~127901:1--4, 2002.

\bibitem{Ekert91}
A.~K. Ekert, ``Quantum cryptography based on \uppercase{B}ell's theorem,'' {\em
  Phys.\ Rev.\ Lett.}, vol.~67, pp.~661--663, 1991.

\bibitem{biham}
E.~Biham, M.~Boyer, P.~O. Boykin, T.~Mor, and V.~Roychowdhury, ``A proof of the
  security of quantum key distribution,'' in {\em Proceedings of the 32nd
  Annual ACM Symposium on Theory of Computing (STOC2000)}, (New York),
  pp.~715--724, ACM Press, 2000.

\bibitem{bdsw}
C.~H. Bennett, D.~A. DiVincenzo, J.~A. Smolin, and W.~K. Wootters,
  ``Mixed-state entanglement and quantum error correction,'' {\em Phys.\ Rev.\
  A}, vol.~54, pp.~3824--3851, 1996.

\bibitem{shorpre}
P.~W. Shor and J.~Preskill, ``Simple proof of security of the \uppercase{BB84}
  quantum key distribution protocol,'' {\em Phys.\ Rev.\ Lett.}, vol.~85,
  pp.~441--444, 2000.

\bibitem{sixstateproof}
H.-K. Lo, ``Proof of unconditional security of six-state quantum key
  distribution scheme,'' {\em Quant.\ Inform.\ and\ Comp.}, vol.~1, no.~2,
  pp.~81--94, 2001.

\bibitem{qkd2waylocc}
D.~Gottesman and H.-K. Lo, ``Proof of security of quantum key distribution with
  two-way classical communications,'' {\em IEEE\ Trans.\ Inf.\ Theo.}, vol.~49,
  pp.~457--475, 2003.

\bibitem{sixstateexact}
H.~F. Chau, ``Practical scheme to share a secret key through a quantum channel
  with a 27.5\% bit error rate,'' {\em Phys.\ Rev.\ A}, vol.~66,
  pp.~060302(R):1--4, 2002.

\bibitem{knill}
A.~Ashikhmin and E.~Knill, ``Non-binary quantum stabilizer codes,'' {\em IEEE\
  Trans.\ Inf.\ Theo.}, vol.~47, pp.~3065--3072, 2001.

\bibitem{sl2q_generators}
A.~A. Albert and J.~Thompson, ``Two-element generation of the projective
  unimodular group,'' {\em Illinois\ J.\ Math.}, vol.~3, pp.~421--439, 1959.

\bibitem{genpur}
G.~Alber, A.~Delgado, N.~Gisin, and I.~Jex, ``Efficient bipartite quantum state
  purification in arbitrary dimensional \uppercase{H}ilbert spaces,'' {\em J.\
  Phys.:A}, vol.~34, pp.~8821--8833, 2001.

\bibitem{multidist}
N.~L. Johnson, S.~Kotz, and N.~Balakrishnan, {\em Discrete Multivariate
  Distributions}.
\newblock New York: Wiley, 1997.
\newblock chap.~39.

\bibitem{hd_ftqc}
D.~Gottesman, ``Fault-tolerant quantum computation with higher-dimensional
  systems,'' {\em Chaos,\ Solitons\ \&\ Fractals}, vol.~10, pp.~1749--1758,
  1999.

\bibitem{Roman}
S.~Roman, {\em Coding And Information Theory}.
\newblock Berlin: Springer, 1992.
\newblock p.~26.

\bibitem{mub1}
W.~K. Wootters and B.~D. Fields, ``Optimal state-determination by mutually
  unbiased measurements,'' {\em Ann.\ Phys.}, vol.~191, pp.~363--381, 1989.

\bibitem{mub2}
J.~Lawrence, C.~Brukner, and A.~Zeilinger, ``Mutually unbiased binary
  observable sets on $\uppercase{N}$ qubits,'' {\em Phys.\ Rev.\ A}, vol.~65,
  pp.~032320:1--5, 2002.

\bibitem{mub3}
S.~Bandyopadhyay, P.~O. Boykin, V.~Roychowdhury, and F.~Vatan, ``A new proof
  for the existence of mutually unbiased bases,'' {\em Algorithmica}, vol.~34,
  pp.~512--528, 2002.

\bibitem{Dickson}
L.~E. Dickson, {\em Linear Groups: With An Exposition Of The Galois Field
  Theory}.
\newblock New York: reprinted edition by Dover, 1958.
\newblock \S260.

\bibitem{Huppert}
B.~Huppert, {\em Endliche Gruppen~I}.
\newblock Berlin: Springer, 1967.
\newblock p.~198 and pp.~213--214.

\bibitem{Hungerford}
T.~W. Hungerford, {\em Algebra}.
\newblock Berlin: Springer, 1974.
\newblock pp.~97--98.

\end{thebibliography}
\end{document}